\documentclass[11pt]{article}
\usepackage{amsmath,amsfonts, float,amssymb,graphicx,geometry,authblk,color,soul,comment,bm,physics,subcaption,siunitx,xcolor} 
\usepackage[colorlinks=true,linkcolor=black,anchorcolor=black,citecolor=black,filecolor=black,menucolor=black,runcolor=black,urlcolor=black]{hyperref} 
\newcolumntype{C}[1]{>{\centering\arraybackslash}p{#1}}
\usepackage{soul}
\usepackage{cleveref}

\title{Characterization of the soft behavior of nematic elastomers over a range of temperature and strain rates}
\author[1]{Alice Kutsyy}
\author[1]{Adeline Wihardja\footnote{Corresponding Author.  Email: awihardja@caltech.edu}}
\author[2]{Victoria Lee}
\author[1]{Kaushik Bhattacharya}
\affil[1]{California Institute of Technology, Pasadena CA 91125}
\affil[2]{Saint-Gobain Research North America, Northborough MA 01532  \footnote{Victoria Lee conducted this work while affiliated with the California Institute of Technology.}}
\date{}

\begin{document}
\graphicspath{images_dt}

\maketitle
\begin{abstract}
     Nematic elastomers are a particular class of liquid crystal elastomers (LCEs) that exhibit both liquid-crystalline order and rubber (entropic) elasticity.  This combination makes them stimuli-responsive soft materials with a number of unusual thermo-mechanical properties.  They have been proposed for various applications, including soft robotics, enhanced adhesion, and impact resistance.  This paper presents a new experimental setup and a comprehensive dataset characterizing the soft behavior of nematic elastomers over a range of temperatures and strain rates.  We also fit the results to a recently developed model of nematic elastomers \cite{lee_2023}. 
\end{abstract}

\section{Introduction} \label{sec:intro}
\begin{figure}
   \centering
   \begin{subfigure}{0.45\textwidth}
     \includegraphics[width=\textwidth]{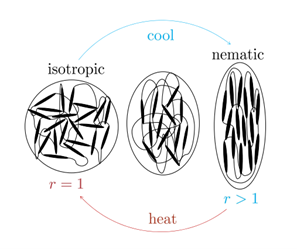}
     \caption{} \label{fig:Mesogens_a}
  \end{subfigure}%
  \hfill
  \begin{subfigure}{0.45\textwidth}
     \includegraphics[width=\textwidth]{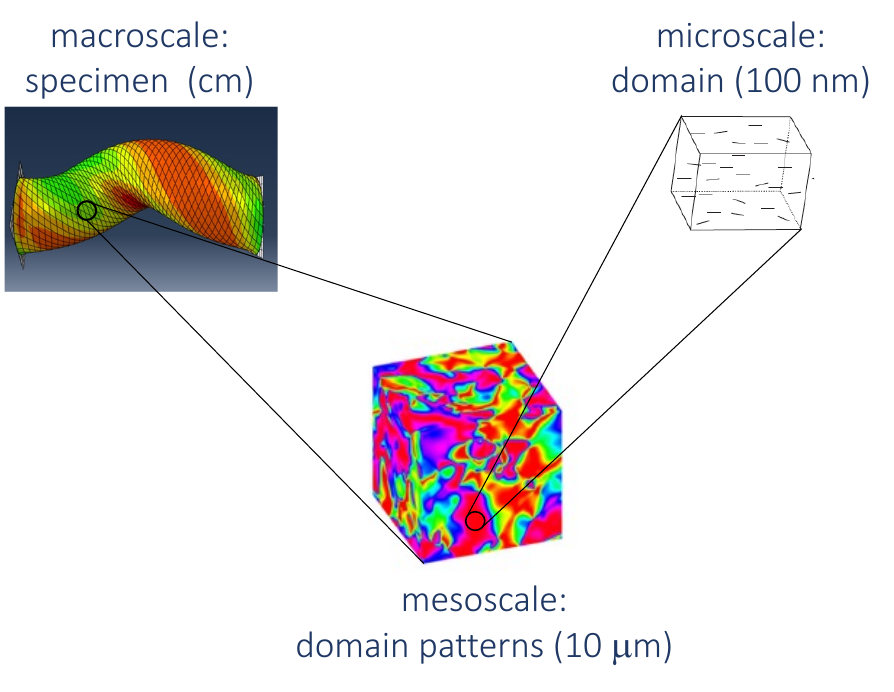}
     \caption{} \label{fig:Mesogens_b}
  \end{subfigure}%
\caption{(a) The temperature-driven isotropic to nematic transition and the resulting deformation in nematic elastomers.  (b) The multiscale nature of I-PLCEs; a typical specimen consists of numerous domains of varying nematic orientation.} \label{fig:Mesogens}
\end{figure}

Liquid crystal elastomers (LCEs) are a class of stimuli-responsive soft materials that exhibit both liquid crystalline order and rubber (entropic) elasticity.   They are composed of lightly cross-linked polymer chains that have nematic mesogens incorporated into their main or side chains \cite{warner_book}.  They can display nematic, smectic, and cholesteric ordering depending on the temperature, material, and crosslinking.  In this work, we consider nematic elastomers: at high temperatures, mesogens are randomly aligned due to entropy, but undergo an isotropic to nematic transition on cooling, and display nematic ordering where the mesogens are aligned in one direction, as shown in Figure \ref{fig:Mesogens}(a).  In particular, we consider isotropic-genesis polydomain nematic LCEs (I-PLCEs) that are synthesized in the isotropic state, but then cooled to a polydomain state where different domains show different nematic ordering directions as shown in Figure \ref{fig:Mesogens}(b).

The combination of entropic elasticity and nematic ordering leads to unusual thermo-mechanical properties in LCEs, and specifically in I-PLCEs.  They can also be activated by chemical, optical, electrical, and magnetic stimuli by incorporating appropriate stimuli-responsive molecules that couple to the nematic ordering.  Therefore, they have been proposed for a variety of applications, including artificial muscles, soft robotics, drug delivery, stretchable membranes, and damping (see \cite{dejeu_2012,ula_2018,herbert_2022,terentjev_2025} for reviews).

\begin{figure}
    \centering
    \includegraphics[width=3.5in]{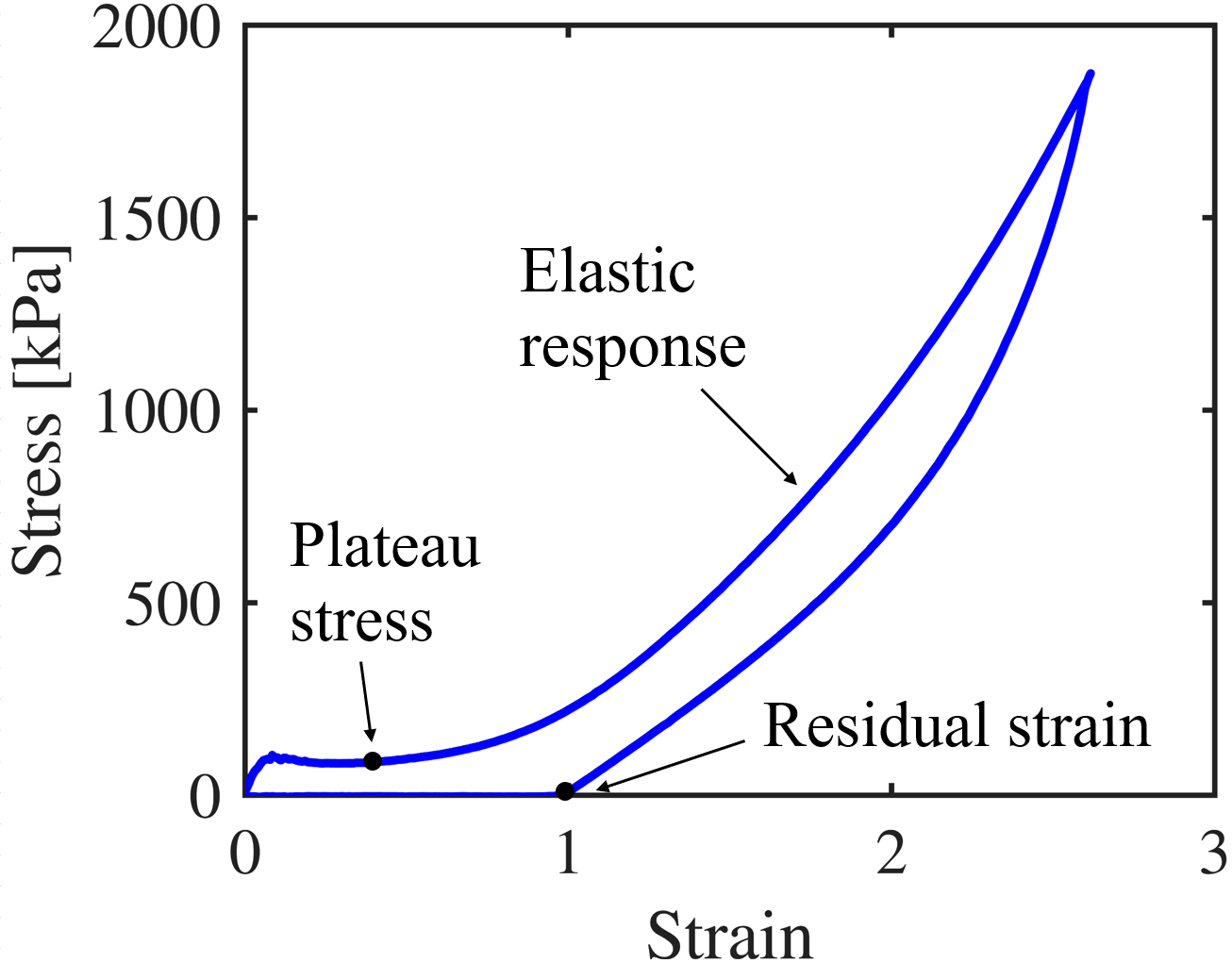}
    \caption{Stress-strain curve demonstrating LCE soft behavior. Plateau stress, elastic response, and residual strain are labeled.}\label{fig:Terms}
    \index{figures}
\end{figure}

An important attribute of I-PLCEs is the soft elastic behavior: the material stretches at almost zero stress, resulting in a soft plateau region in the stress-strain curve \cite{clarke1998texture,biggins_warner_bhatta,urayama_kohmon,zhou} as shown in Figure \ref{fig:Terms}. This is attributed to the polydomain-monodomain transition (PMT): the material, as synthesized, is polydomain, but the mesogens rotate and the domain patterns change until all the domains and mesogens are aligned with the principal stress.  This PMT-induced soft behavior results in reduced wrinkling in thin sheets \cite{plucinsky2017microstructure}, high energy absorption in impact \cite{jeon2022synergistic}, surface instabilities \cite{biggins_2023}, in-plane liquid-like phenomena \cite{tokumoto_zhou}, anomalous behavior in contact \cite{maghsoodi_2023}, and enhanced adhesion \cite{farre2022dynamic,ohzono2019enhanced,maghsoodi_2025}.   The complex behavior of I-PLCEs has been described in a constitutive law by Lee {\it et al.} \cite{lee_2023}, which has been used in finite element analysis to study in-plane liquid-like behavior \cite{lee_2023}, contact \cite{maghsoodi_2023}, adhesion \cite{maghsoodi_2025}, amongst others.  

There has been considerable effort to characterize the thermo-mechanical behavior and properties of polydomain LCEs, both I-PLCEs synthesized in the isotropic state and N-PLCEs synthesized in the nematic state. The behavior of N-PLCEs at various strain rates at room temperature, and across temperatures ranging from room temperature to the transition temperature at a low strain rate is examined in Azoug {\it et al.} \cite{azoug_2016}. In this work, however, the soft-elasticity behavior in N-PLCEs is limited by the polymer network's entropic constraints. Still, they exhibit a PMT regime, which is found to depend on both strain rate and temperature. I-PLCEs with low cross-link density have been studied in Linares {\it et al.} \cite{linares_2020}, demonstrating the characteristic soft-elasticity region at different strain rates and mesogen alignment order parameters. However, temperature effects are not probed. This work found that the plateau stress and soft-elasticity regime vary strongly with strain rate, over the range 10$^{-2}$ to 10 s$^{-1}$. In both of these studies, however, the cross-link density is kept fixed. Recently,  Traugutt {\it et al.} \cite{Traugutt} examined both I-PLCEs and N-PLCEs across multiple cross-link densities, showing that the PMT regime and soft behavior in both types are significantly affected by cross-link density. All of the above work concerns main-chain PLCEs synthesized via the Michael addition click reaction as described in \cite{yakacki_RSC}. Other studies by Terentjev and collaborators \cite{hotta_terentjev,clarke_terentjev} have explored side-chain polydomain nematic elastomers and their temperature dependence from room temperature to the phase transition temperature at a low strain rate of 10$^{-4}$ s$^{-1}$ \cite{hotta_terentjev} and strain-rate dependence at low rates (10$^{-4}$ to 10$^{-7}$ s$^{-1}$) \cite{hotta_terentjev,clarke_terentjev}.

While these works have provided significant insight and have motivated comprehensive constitutive models \cite{lee_2023,wang,zhou_2025}, the increasing engineering applications \cite{dejeu_2012,ula_2018,herbert_2022,terentjev_2025} and the complexity of these constitutive models call for a comprehensive dataset over a range of temperatures, strain rates, and cross-link density.  This motivates the current work.  We describe a new experimental setup and a comprehensive dataset characterizing the soft behavior of isotropic-genesis polydomain nematic elastomers over a range of temperatures and strain rates. We also fit the results to a recently developed model of nematic elastomers \cite{lee_2023}.

%
%
%

\section{Experimental methods} \label{sec:exp}

\subsection {Experimental setup} \label{sec:design}

\begin{figure}
    \centering
   \begin{subfigure}{0.9\textwidth}
     \includegraphics[width=\linewidth]{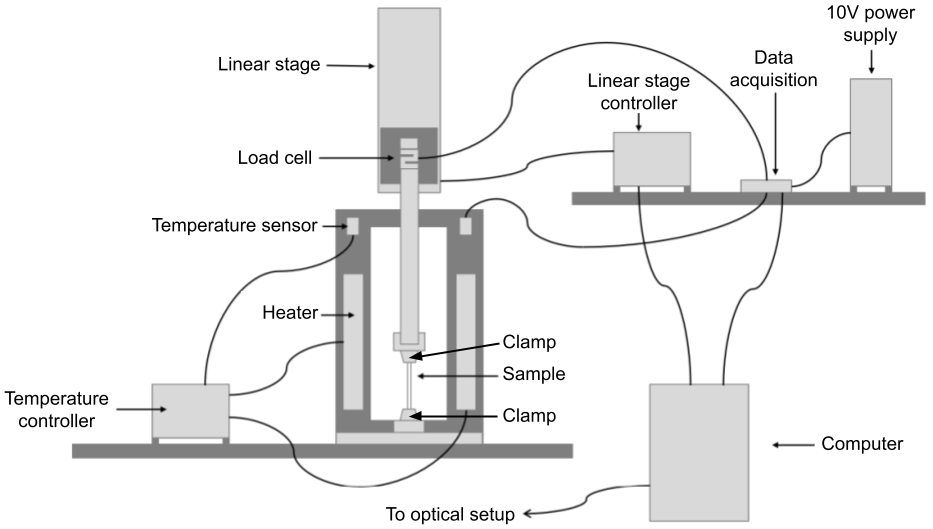}
    \caption{}\label{fig:Small_schematic}
     \end{subfigure}\\
   \begin{subfigure}{0.6\textwidth}
     \includegraphics[width=\linewidth]{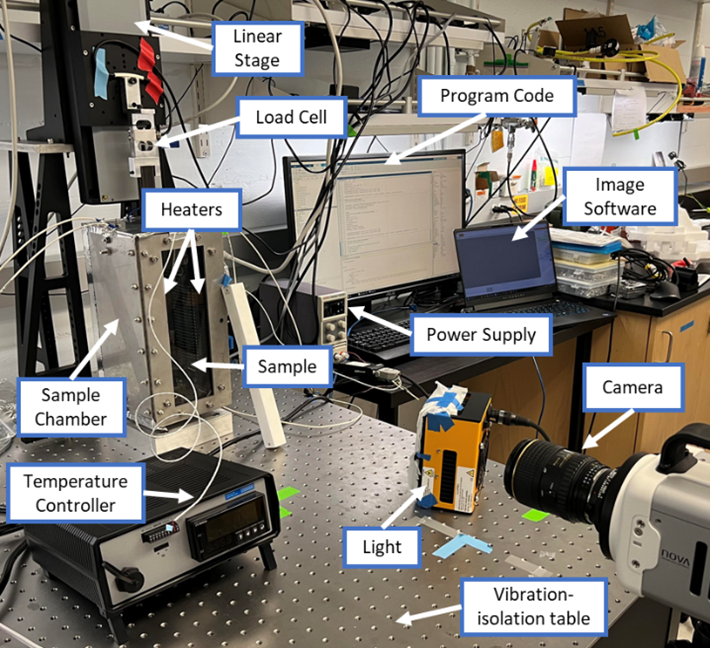}
     \caption{} \label{fig:Rig_photo_a}
  \end{subfigure}%
  \hfill
  \begin{subfigure}{0.25\textwidth}
     \includegraphics[width=\linewidth]{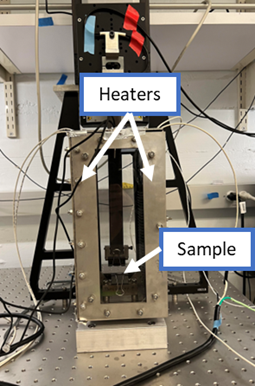}
     \caption{} \label{fig:Rig_photo_b}
  \end{subfigure}%
  \hfill
    \begin{subfigure}{0.1\textwidth}
     \includegraphics[width=\linewidth]{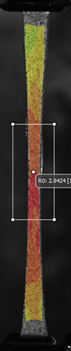}
     \caption{} \label{fig:DIC}
  \end{subfigure}%
\caption{The experimental setup and typical result.  (a) Schematic diagram, (b) the overall setup  (c) the heated chamber, and (d) a typical DIC result.} \label{fig:Rig_photo}
\end{figure}

The experimental setup used to perform tensile tests at various strain rates, temperatures, and cross-linker densities is shown in Figure \ref{fig:Rig_photo}.   It consists of a heated chamber with a window, a linear translation stage, an imaging system, various controllers, and a data acquisition system.

The chamber (see Figure \ref{fig:Rig_photo_b}) is insulated on the sides and equipped with heaters and resistance temperature detector (RTD) sensors. The temperature of the chamber is controlled using an Omega CS8DPT Benchtop controller.  The maximum deviation over all tests is 8$^\circ$C, and occurred during the slowest strain rate trial ($10^{-3}$ s$^{-1}$) at 90\si{\degreeCelsius}; typical deviations are much smaller.

The chamber contains a stationary bottom clamp.  The specimen is held in the chamber between the stationary bottom clamp and a top clamp that is attached to a pull-rod that emerges from the chamber.  The pull-rod is attached to a Physik Instrumente linear stage and controller, and an Omega LC101-50 load cell.  

The specimen is illuminated and imaged through the window using a Photron FASTCAM NOVA S12 high-speed camera equipped with a 100 mm Tokina AT-X Pro lens at 2 frames per second and a resolution of 1024 by 1024 pixels for the duration of the test. The images are analyzed using 2D digital image correlation (DIC) (Vic2D, Correlated Solutions, Columbia, SC) with a typical 25 pixel subset size and step size of 1 to obtain the displacement field. A typical strain map is shown in Figure \ref{fig:DIC}.  We observe that the deformation is largely uniform across the specimen with no stripes or other almost periodic oscillations.  There is some non-uniform strain near the grips due to the shape of the specimen and the effect of grips.  We find good correlation for the 50 mol\% specimens, but not the 25 mol\% specimens, due to the large strains in the latter. Therefore, we use the imaged strain in the gage section for the 50 mol\% specimens, but calculated strain from the clamp-to-clamp distance in the 25 mol\% specimens.
The stress and strain values are correlated using their first respective non-zero value.

\subsection {Liquid Crystal Elastomer Synthesis} \label{sec:synthesis}
We study polydomain, isotropic-genesis LCEs (I-PLCEs) fabricated using a Michael addition click reaction for main-chain LCEs according to Saed {\it et al.} \cite{JOVE}. We synthesize I-PLCEs with 50 mol\% and  25 mol\% cross-linker density (this indicates molar ratio of thiol functional groups between PETMP and EDDET).   We focus on these higher concentrations to minimize the network viscosity and focus on the soft behavior due to domain reorientation, consistent with previous studies \cite{JOVE,tokumoto_zhou,Traugutt}. Others have used lower densities to probe network viscosity \cite{azoug_2016}, or used second stage crosslinking \cite{linares_2020}.
Toluene is used to dissolve the di-acrylate mesogens (RM 257) at 80\si{\degreeCelsius} \ until a clear solution emerges, and then the spacer (2,2-(ethylenedioxy) diethanethiol) and crosslinker (pentaerythritol tetrakis) are added. The catalyst solution (dipropylamine) diluted in Toluene is then added to the mixture. The solution is mixed and vacuumed to remove any air bubbles. The resulting solution is poured into molds and cured at room temperature before being heated in an oven at 80\si{\degreeCelsius} under a 71.1 kPa vacuum for 12 hours to complete the polymerization at the isotropic phase. Samples are cooled, unmolded, and then heated briefly to remove any stretch. All tests are performed within 2-14 hours after samples are fully cured. The length of the tested specimens is 0.05 m, with a thickness of 0.0015 m. The width of the specimen varies from 0.012 m to 0.005 m in the gauge section. We study the resulting I-PLCE from this method with differential scanning calorimetry (DSC)  and recover a nematic to isotropic phase transition temperature ($T_{ni}$) of 87 \si{\degreeCelsius}.




\section{Experimental observations}\label{sec:results}

We probe the behavior of I-PLCEs at four strain rates (5$\times 10^{-2}$, $10^{-2}$, 5$\times 10^{-3}$, $10^{-3}$ s$^{-1}$), three temperatures (26\si{\degreeCelsius} or room, 55\si{\degreeCelsius} or intermediate, 90\si{\degreeCelsius} or high), and two cross-linker densities (50 mol\%, 25 mol\%).   
The nematic to isotropic transformation temperature $T_\text{ni}$ for the specimen is measured to be 87\si{\degreeCelsius} using differential scanning calorimetry.  Therefore, two of the temperatures are below the nematic-isotropic transition temperature $T_\text{ni}$, and one is above it. Unless otherwise noted, all samples in each figure are from the same synthesis batch, and all strains are Lagrangian strains.

\subsection{Tests on specimens with cross-linker density 50 mol\%} \label{sec:50mol}


\paragraph{Tensile tests at room temperature (26\si{\degreeCelsius})} \label{sec:50roomtemp}

\begin{figure}[t]
   \centering
   \begin{subfigure}{0.48\textwidth}
     \includegraphics[width=\linewidth]{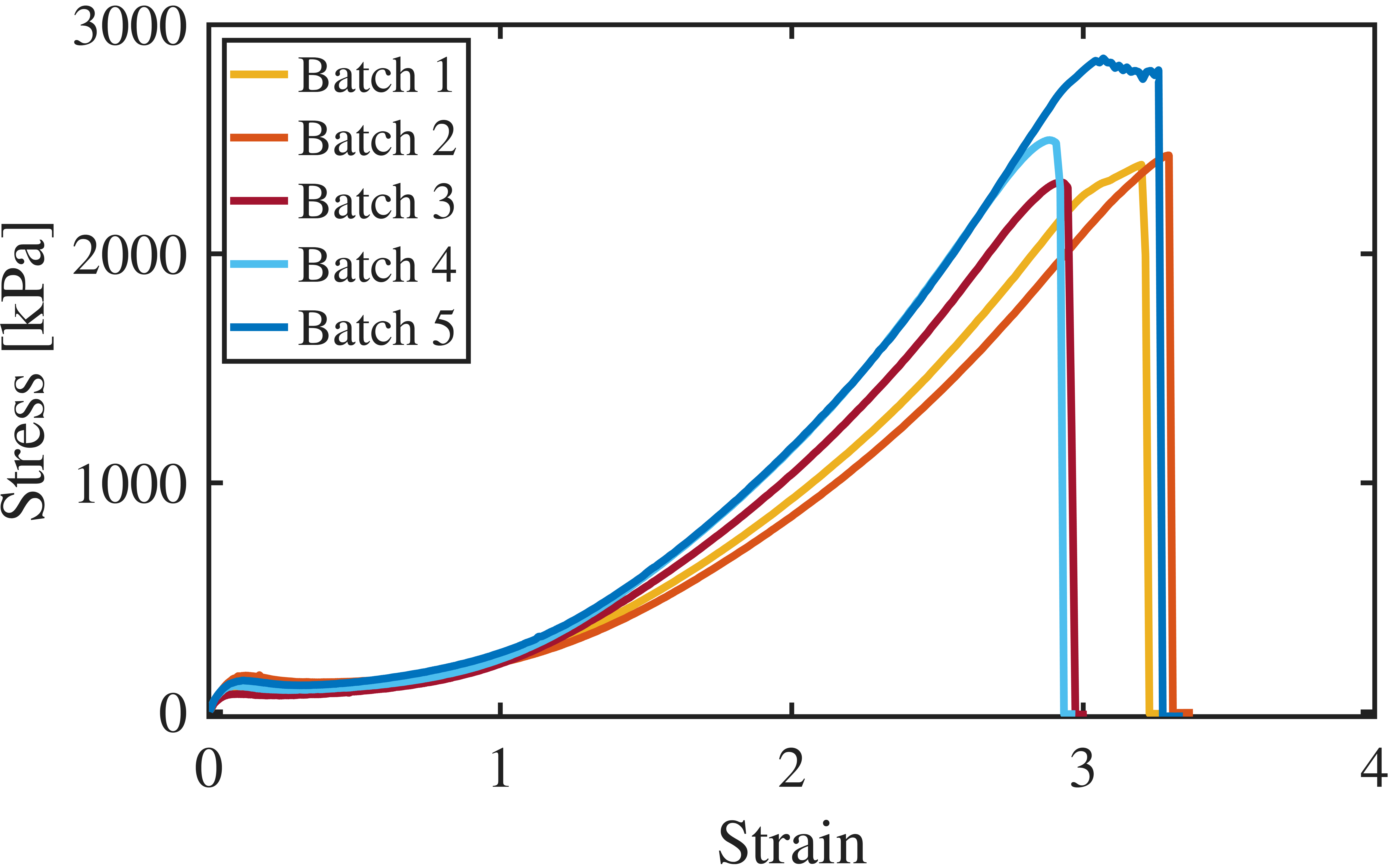}
     \caption{} \label{fig:Batches}
  \end{subfigure}%
  \hspace{0.01\textwidth}
  \begin{subfigure}{0.48\textwidth}
     \includegraphics[width=\linewidth]{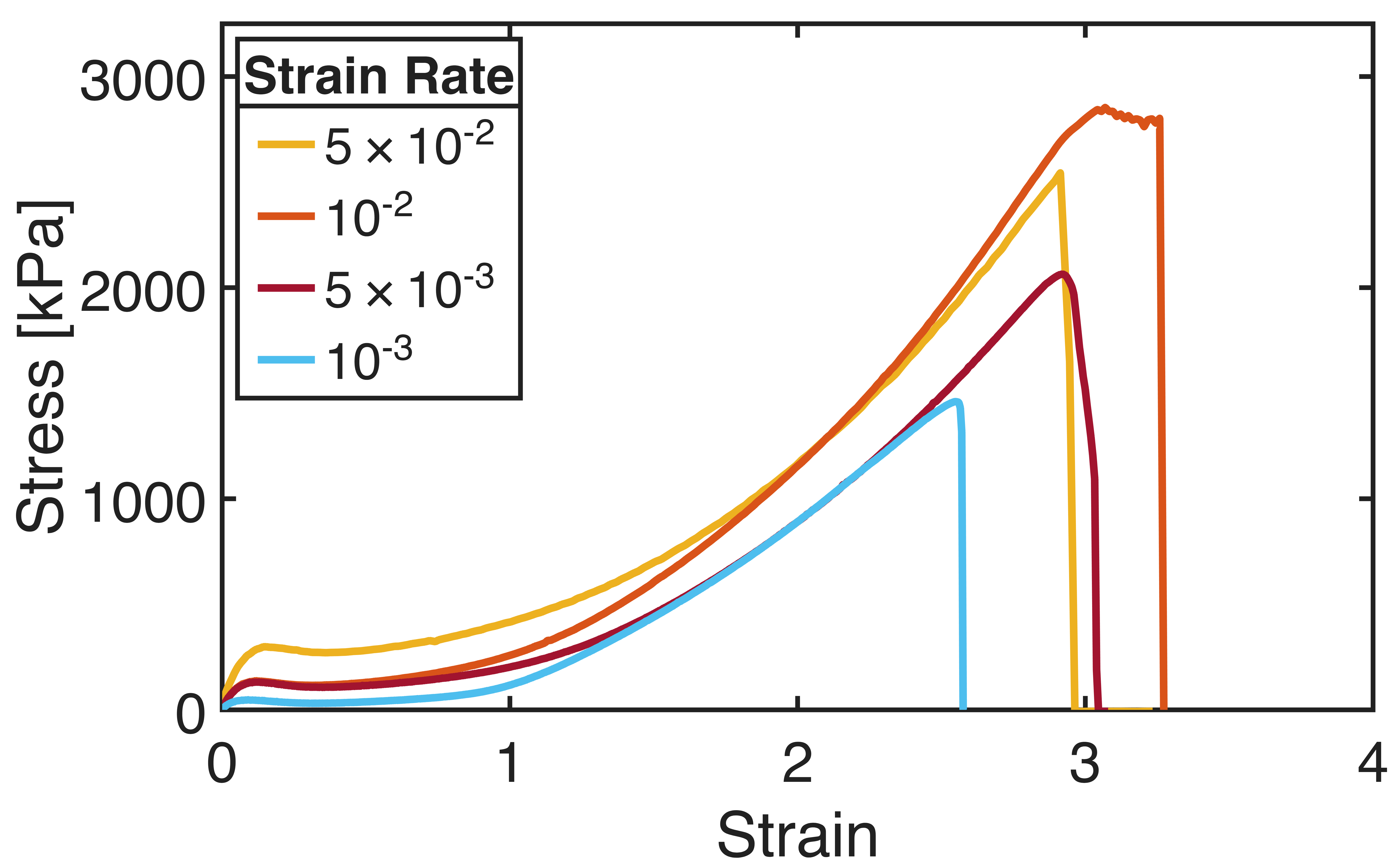}
     \caption{} \label{fig:RT_Failure}
  \end{subfigure}\\
   \begin{subfigure}{0.48\textwidth}
     \includegraphics[width=\linewidth]{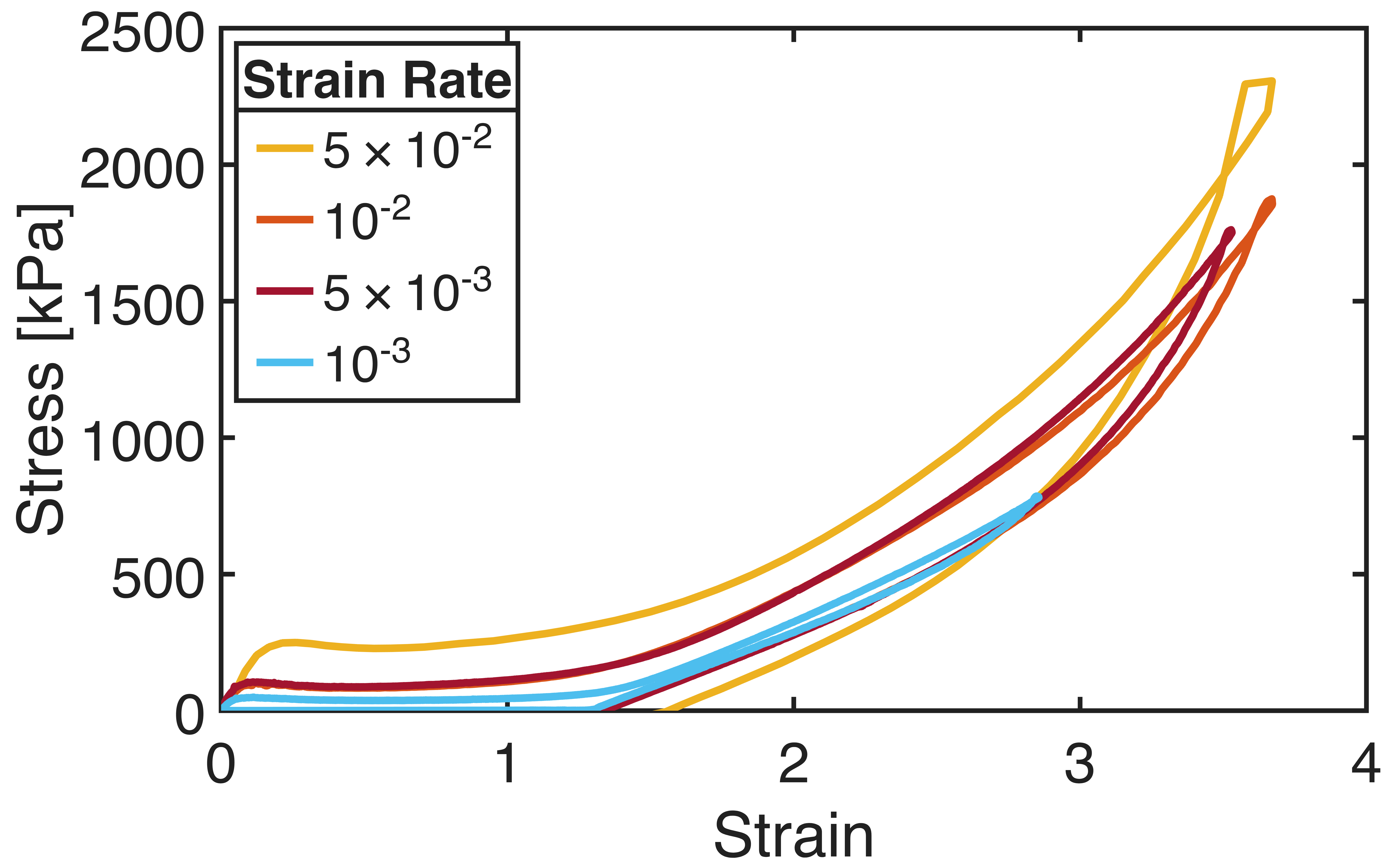}
     \caption{} \label{fig:RT_Load-Unload_a}
  \end{subfigure}%
  \hspace{0.01\textwidth}
  \begin{subfigure}{0.48\textwidth}
     \includegraphics[width=\linewidth]{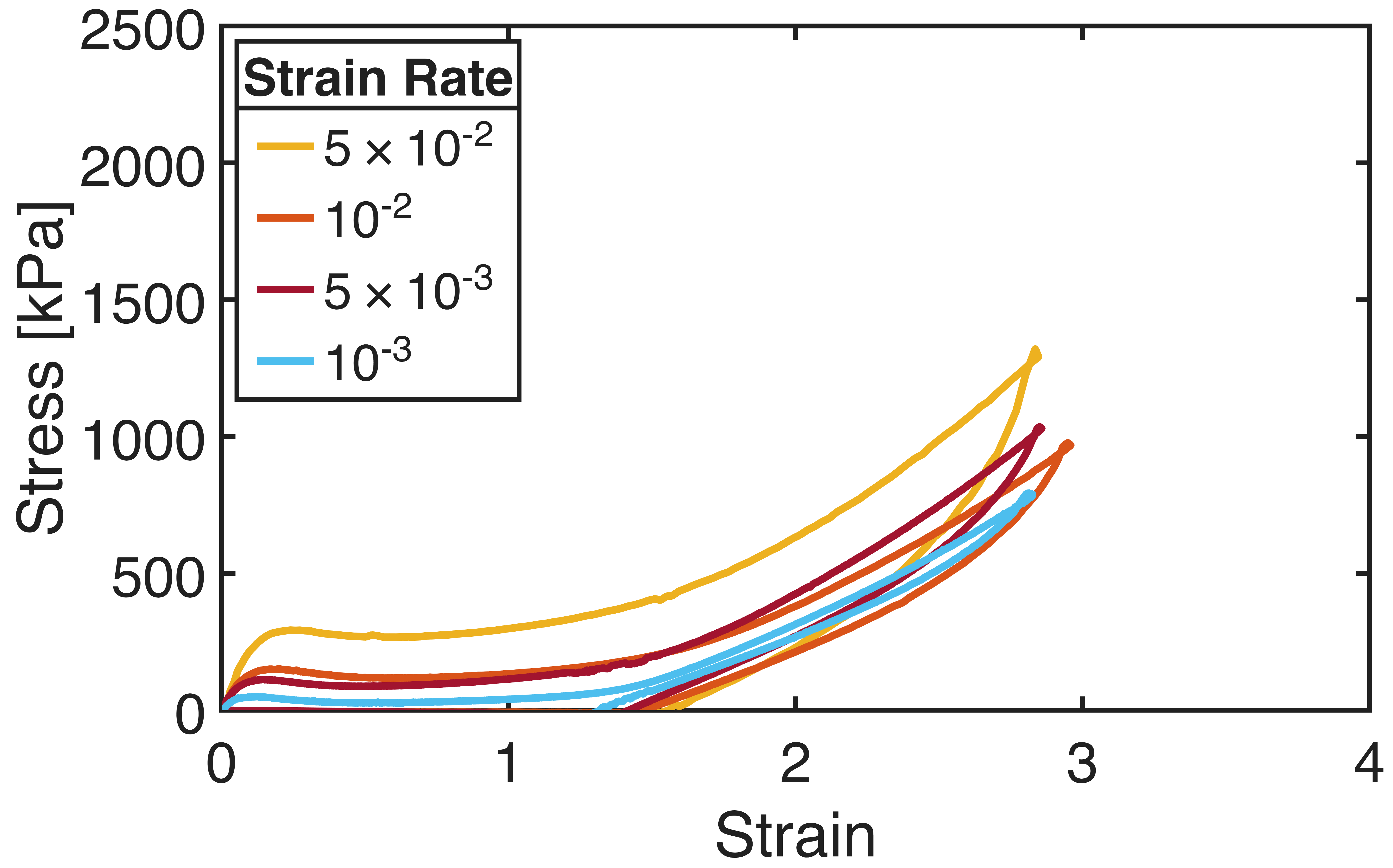}
     \caption{} \label{fig:RT_Load-Unload_b}
  \end{subfigure}%
\caption{Room temperature tensile tests for 50 mol\% cross-linker density. (a) Test to failure for a number of batches, and}  (b)  for various loading rates.  (b) L:oad-unload tests for the first batch tested to a maximum strain of 3.5, and (c) load-unload tests for the second batch tested to a maximum strain of 3.0.
 \label{fig:RT_Fail-Load-Unload}
\end{figure}

\begin{table}
    \centering
    \caption{Maximum nominal (clamp-to-clamp) Lagrangian strains for load-unload curves.}
    \label{tab:strain}
    \renewcommand{\arraystretch}{1.25}
    \begin{tabular}[t]{ >{\raggedright\arraybackslash}p{6em}| C{6em} C{6em} C{6em} C{4em} } 
        \hline
         Strain Rate & $5\times10^{-2}$ s$^{-1}$ & $10^{-2}$ s$^{-1}$ & $5\times10^{-3}$ s$^{-1}$ & $10^{-3}$ s$^{-1}$ \\ 
        \hline
       RT or 26\si{\degreeCelsius} & 2.625, 2.031 & 2.625, 2.031 & 2.625, 2.031 & 2.031 \\ 
        \hline
        55\si{\degreeCelsius} & 0.861 & 0.861 & 0.625 & 0.625 \\ 
        \hline
        90\si{\degreeCelsius} & 0.48 & 0.345 & 0.345 & 0.345  \\ 
        \hline
    \end{tabular}
\end{table}

We first perform tensile test to failure, and the results are shown in  Figures \ref{fig:Batches} and \ref{fig:RT_Failure}.  We observe  in Figure \ref{fig:Batches} that the batch to batch variation is small.  Figure \ref{fig:RT_Failure} shows the rate-dependence, and provides the maximum strain that captures the general behavior without risking failure.  We then conduct load-unload tests for strain up to the maximum values, and the results are shown in Figures \ref{fig:RT_Load-Unload_a} and \ref{fig:RT_Load-Unload_b}.  These are performed for two batches of specimens; each batch is tested at various strain rates, and the two batches are subjected to different maximum stretches.  The overall behavior is consistent across the two batches. 

The observed response clearly shows the well-known soft behavior of I-PLCEs \cite{urayama_kohmon} marked by a notable plateau in the stress-strain response, and a change in opacity of the specimen.  This is a result of the polydomain-monodomain transition (PMT): nematic directors are randomly oriented in the as fabricated polydomain specimens but align along the loading direction, thereby accommodating a larger strain with no increase in stress \cite{urayama_kohmon,biggins_warner_bhatta,zhou}.  This change also manifests as a change of opacity: the sample is opaque due to light scattering of the random directors, but becomes transparent as they align.  The PMT  also manifests as a residual strain, approximately 1.5 across all strain rates and both batches. 

We observe small loops at the end of loading and beginning of unloading in some of our tests.  These are artifacts caused by two reasons.  First, we measure forces using a load cell and strain from raw images processed with DIC.  A synchronization error between the two channels of information can lead to these loops.   Second, we observe that the specimens with loops develop small out-of-plane motion during the early part of the unloading.  There are a few possible reasons: grip effects, some slip at the grips, some temperature gradients at the grips, and the kinetics of reverse transformation.  Since we use 2D-DIC to measure displacements and strains, any out-of-plane motion can lead to artifacts in the correlation.  In particular, the out-of-plane motion would lead to a smaller apparent strain and manifest as loops.


The PMT and the resulting stress plateau depend on the strain rate: at a strain rate of $5\times 10^{-2}$ s$^{-1}$, the stress plateau is around 235 kPa and the soft behavior extends to approximately 0.9 strain, while at $1\times 10^{-3}$ s$^{-1}$, the stress plateau is around 35 kPa and the soft behavior extends to around 1.2 strain.  The lower strain rates offer greater time for the domains and directors, as well as the underlying elastomers, to relax to their preferred orientations \cite{fridrikh_terentjev_1999,clarke_1998}.

We observe elastic behavior with hardening beyond the plateau as expected for elastomers. It is hard to discern any strain rate effects in our observations.

%

\paragraph{Tensile tests at 55\si{\degreeCelsius}} \label{sec:55}

\begin{figure}
   \centering
   \begin{subfigure}{0.48\textwidth}
     \includegraphics[width=\linewidth]{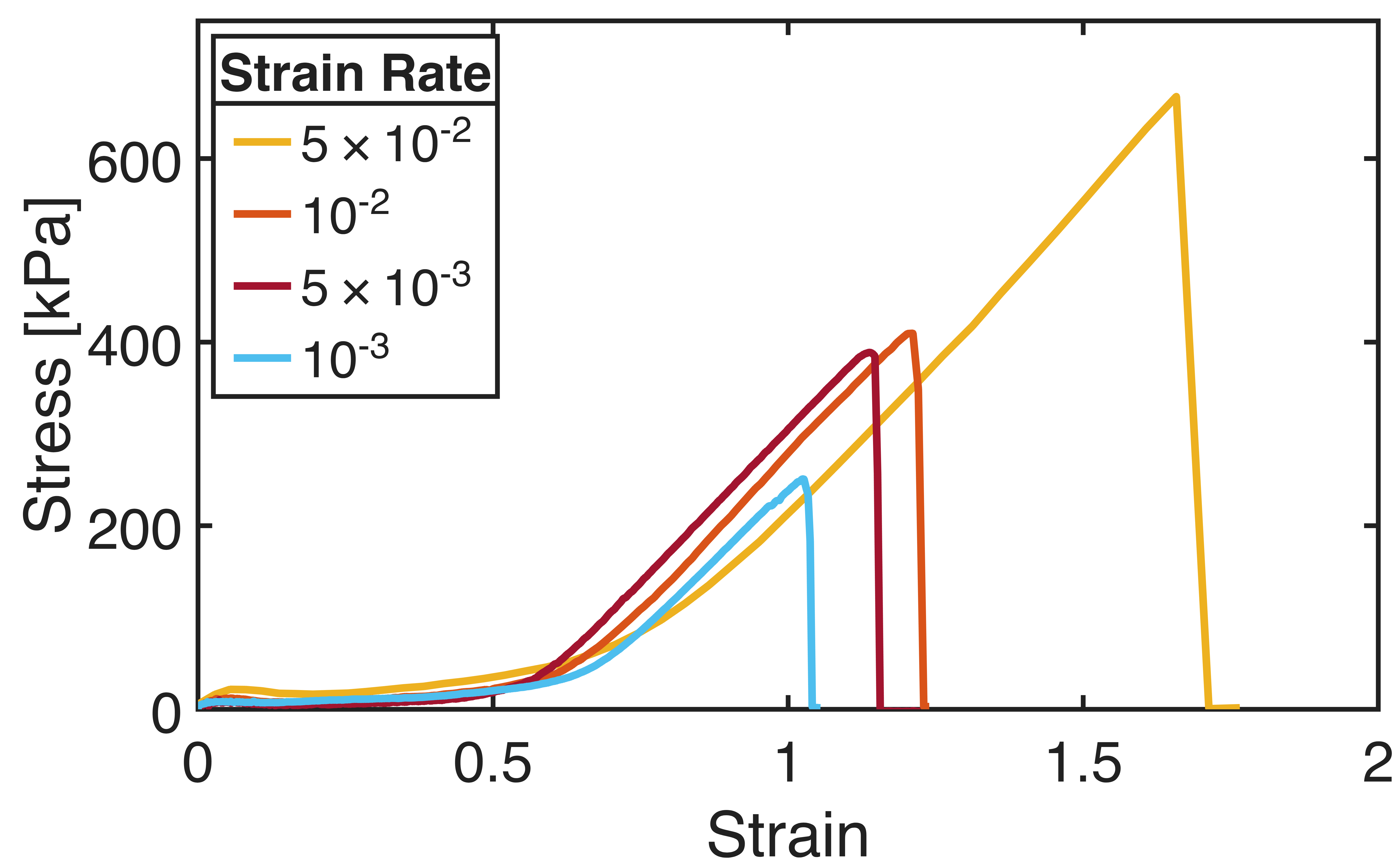}
     \caption{} \label{fig:55_Failure}
  \end{subfigure}%
  \hspace{0.01\textwidth}
  \begin{subfigure}{0.48\textwidth}
     \includegraphics[width=\linewidth]{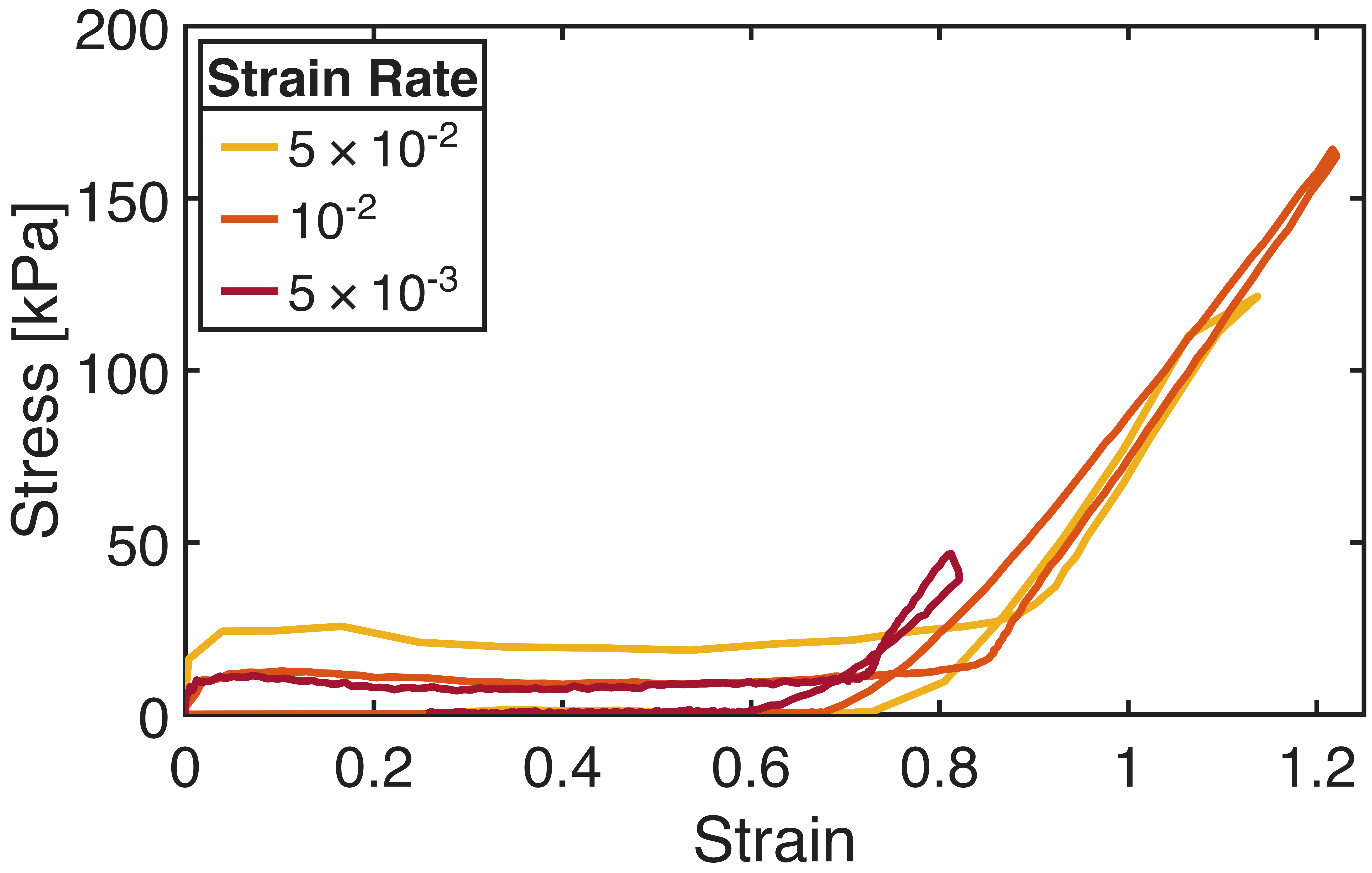}
     \caption{} \label{fig:55_Load-Unload}
  \end{subfigure}%
\caption{55$^\circ$C tensile tests for 50 mol\% cross-linker density at various strain rates. (a) Tests to failure,  and (b) load-unload tests.}\label{fig:55}
\end{figure}

The results of the tests at 55\si{\degreeCelsius} are shown in Figure \ref{fig:55}. The plateau stress increases with strain rate, as seen before, although it is nearly identical between the $10^{-2}$ s$^{-1}$ and 5$\times 10^{-3}$ s$^{-1}$ strain rates. The extent of the plateau regime is approximately 0.8 strain, and the residual strain is approximately 0.7.  As in the case with room temperature tests, we attributed the loop seen in our measurements due to misalignment in synchronization of the stress and strain readings, which were recorded on different systems and synchronized using their first respective non-zero value. Additionally, it may be due to slight out-of-plane deformations, especially near the grips due to temperature gradient.

\paragraph{Tensile tests at 90\si{\degreeCelsius}} \label{sec:90}

\begin{figure}
   \centering
   \begin{subfigure}{0.49\textwidth}
     \includegraphics[width=\linewidth]{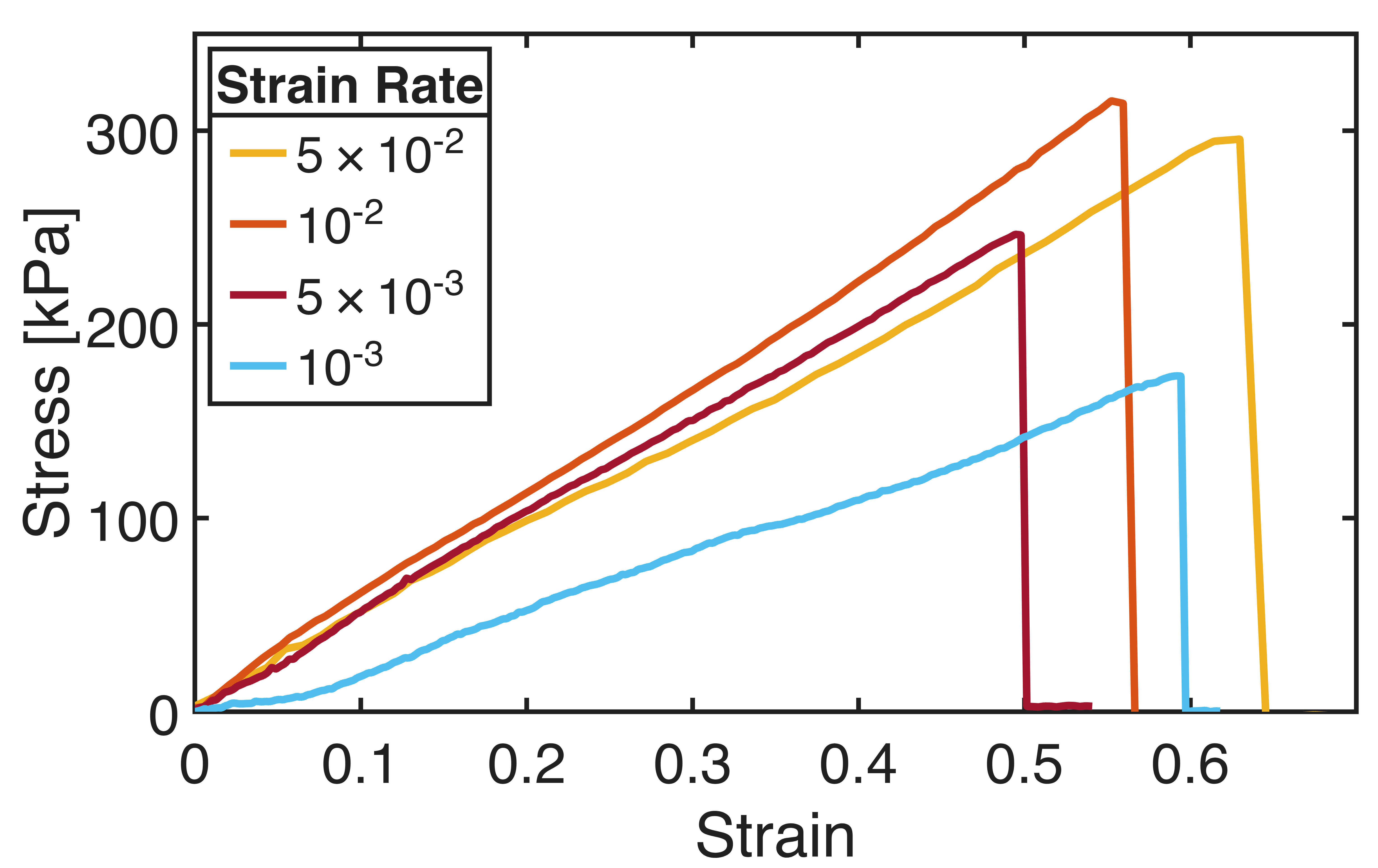}
     \caption{} \label{fig:90_Failure}
  \end{subfigure}%
  \hspace{0.01\textwidth}
  \begin{subfigure}{0.48\textwidth}
     \includegraphics[width=\linewidth]{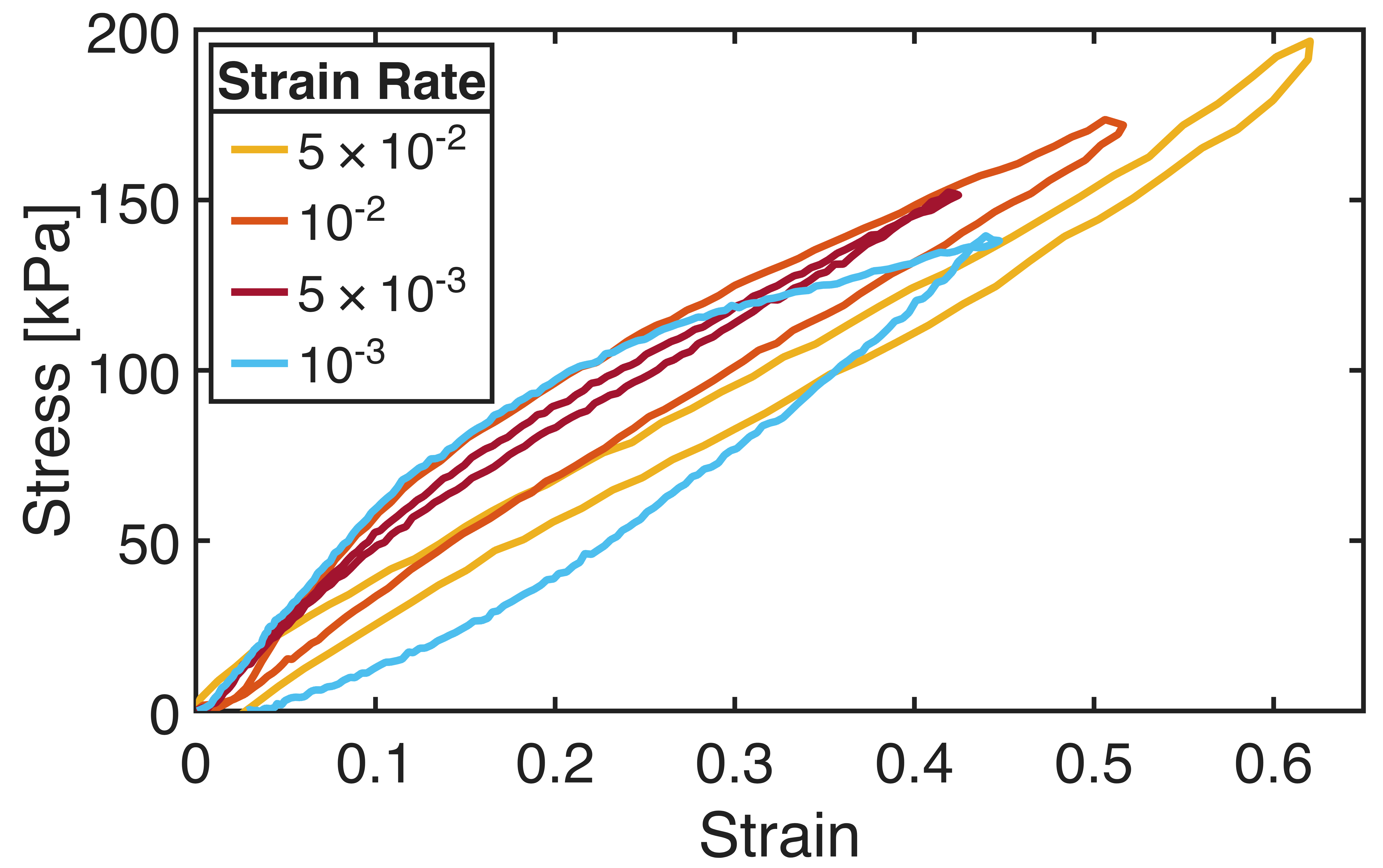}
     \caption{} \label{fig:90_Load-Unload}
  \end{subfigure}%
\caption{90$^\circ$C tensile tests for 50 mol\% cross-linker density at various strain rates. (a) Test to failure,  (b) load-unload tests.}\label{fig:90}
\end{figure}

The results of the tests at 90\si{\degreeCelsius} are shown in Figure \ref{fig:90}.  We do not observe any stress plateau, and there is negligible residual strain.     Since the test temperature is above  $T_\text{ni}$ of 87 \si{\degreeCelsius}, the specimen is in the isotropic state and we do not expect any soft behavior.  There are also no noticeable strain rate effects.

\paragraph{Comparison across temperatures} \label{sec:acrosstemps}

\begin{figure}
   \centering
 \begin{subfigure}{0.8\textwidth}
     \includegraphics[width=\linewidth]{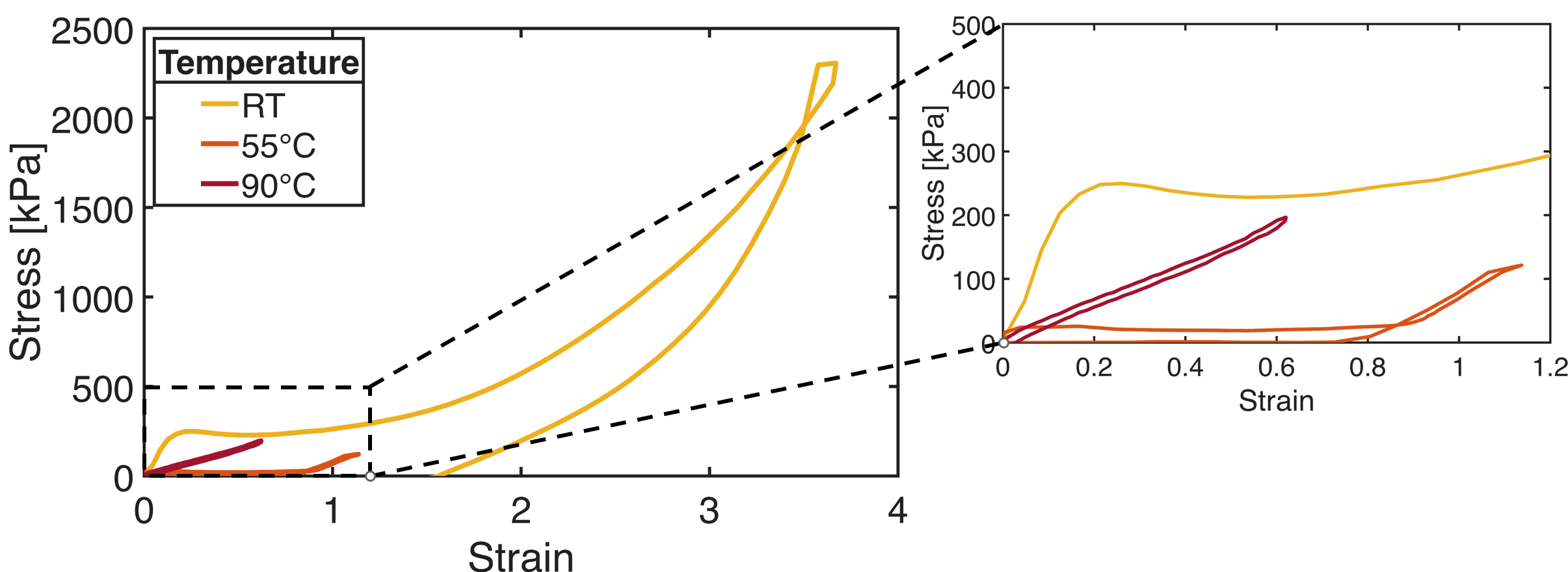}
     \caption{} \label{fig:Across_Temps_a}
  \end{subfigure}%
  \hspace{0.01\textwidth}
  \begin{subfigure}{0.8\textwidth}
     \includegraphics[width=\linewidth]{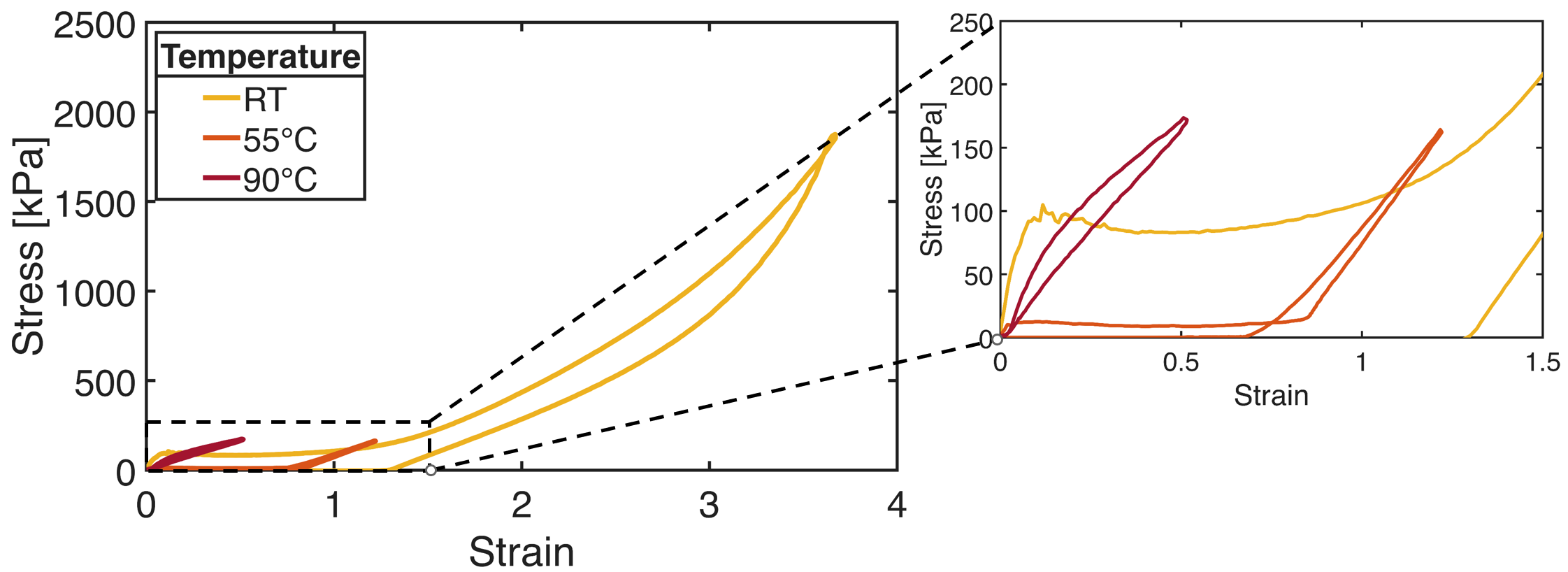}
     \caption{} \label{fig:Across_Temps_b}
  \end{subfigure}%
  \hspace{0.01\textwidth}
  \begin{subfigure}{0.8\textwidth}
     \includegraphics[width=\linewidth]{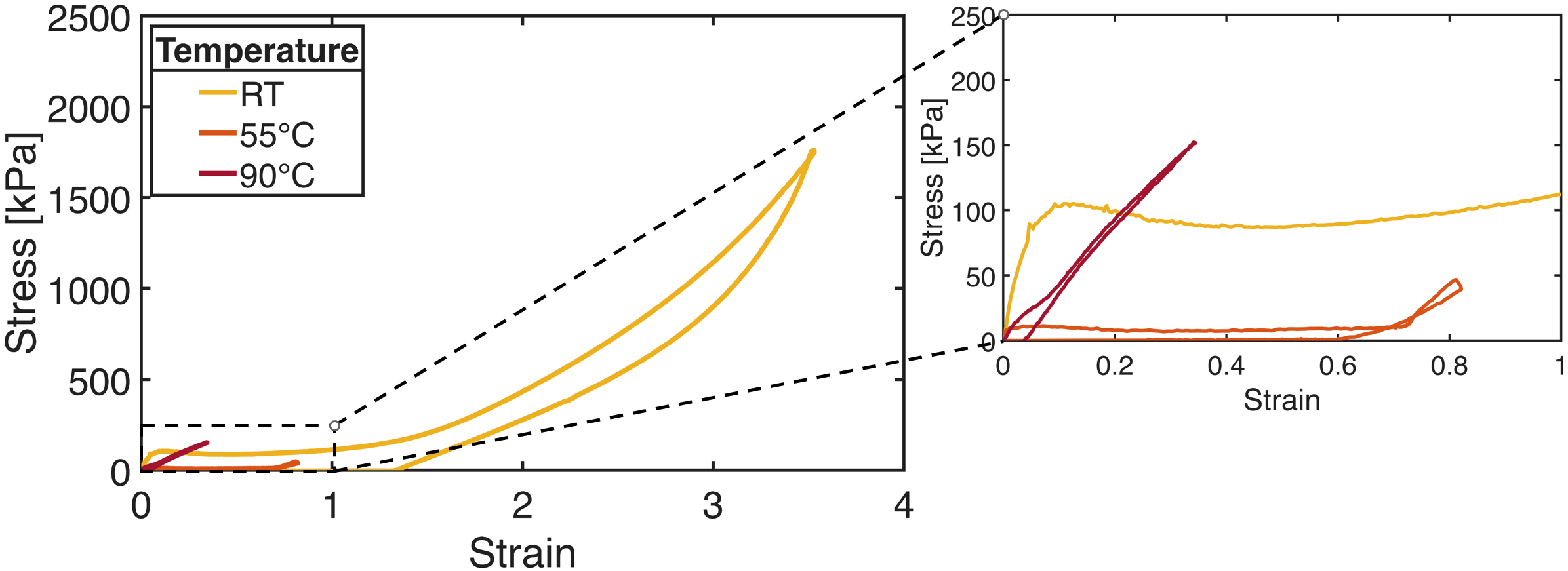}
     \caption{} \label{fig:Across_Temps_c}
  \end{subfigure}%
  \hspace{0.01\textwidth}
  \begin{subfigure}{0.8\textwidth}
     \includegraphics[width=\linewidth]{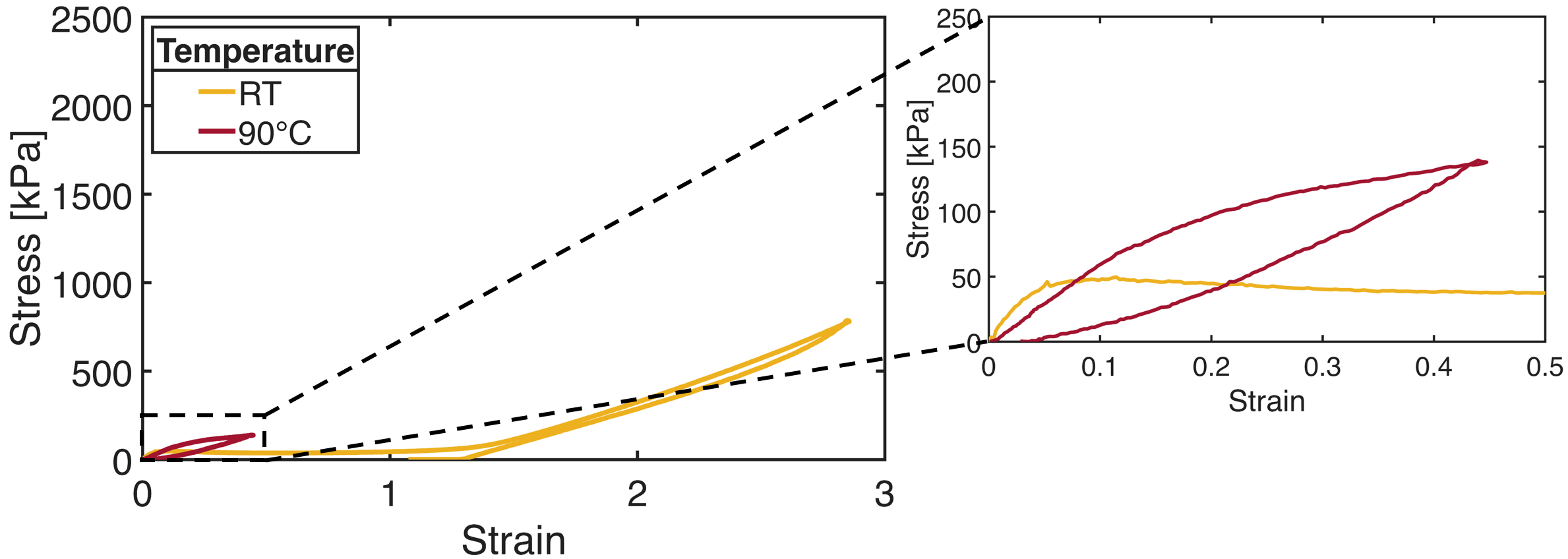}
     \caption{} \label{fig:Across_Temps_d}
  \end{subfigure}%
\caption{Comparison across temperatures for trials with a strain rate of (a) 5$\times 10^{-2}$ s$^{-1}$, (b) $10^{-2}$ s$^{-1}$, (c) 5$\times 10^{-3}$ s$^{-1}$, (d) $10^{-3}$ s$^{-1}$.} \label{fig:Across_Temps}
\end{figure}

We revisit the results above by looking at the effect of temperature at the various strain rates in Figure \ref{fig:Across_Temps}.  While the tests are conducted with different batches at different temperatures, the batch to batch variation as shown in Figure \ref{fig:Batches} is small compared to the temperature variation (also see Appendix \ref{app:var}).  

\begin{figure}
   \centering
 \begin{subfigure}{0.48\textwidth}
     \includegraphics[width=\linewidth]{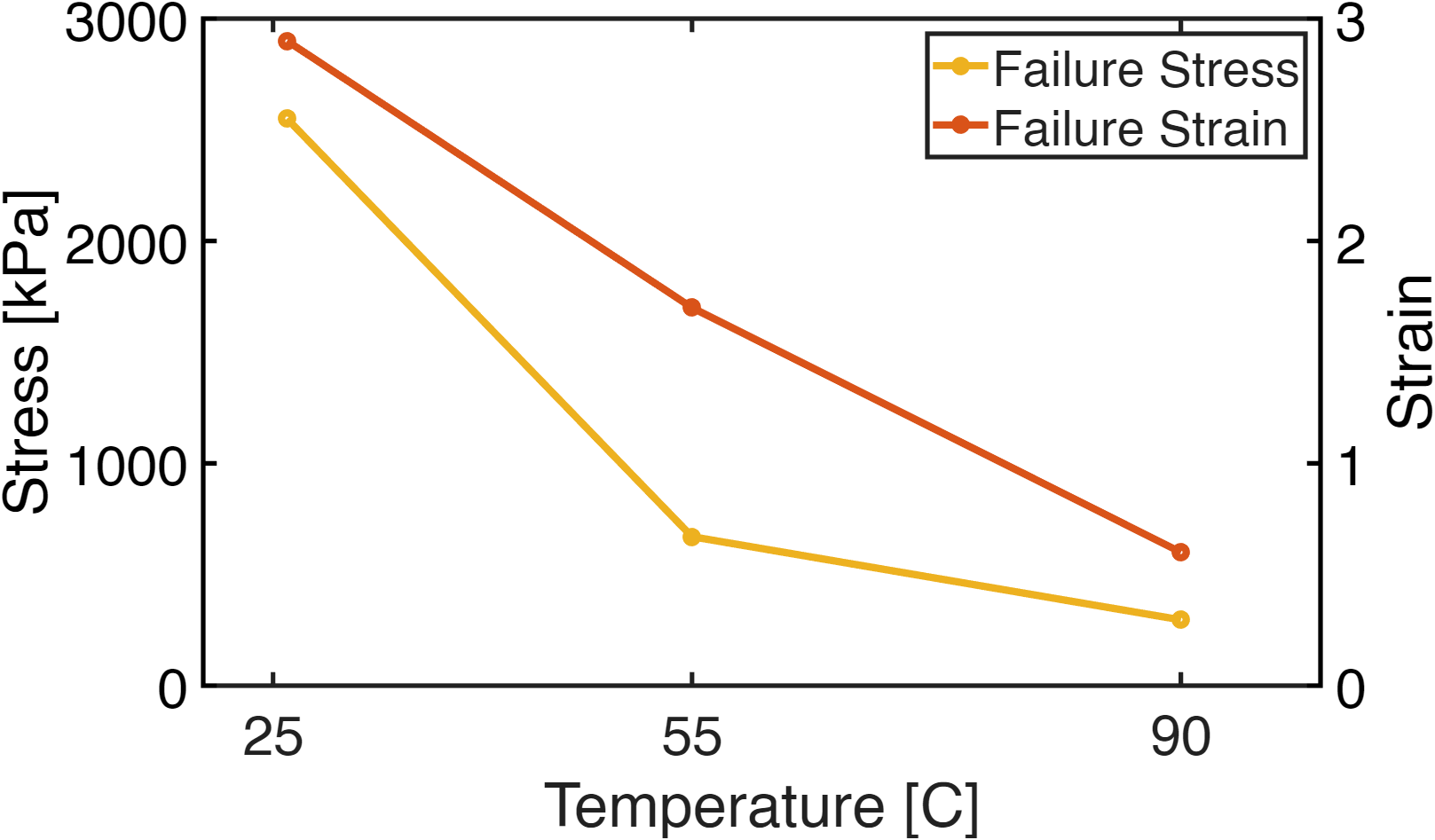}
     \caption{} \label{fig:fail_summ}
  \end{subfigure}
  \hspace{0.01\textwidth}
 \begin{subfigure}{0.48\textwidth}
     \includegraphics[width=\linewidth]{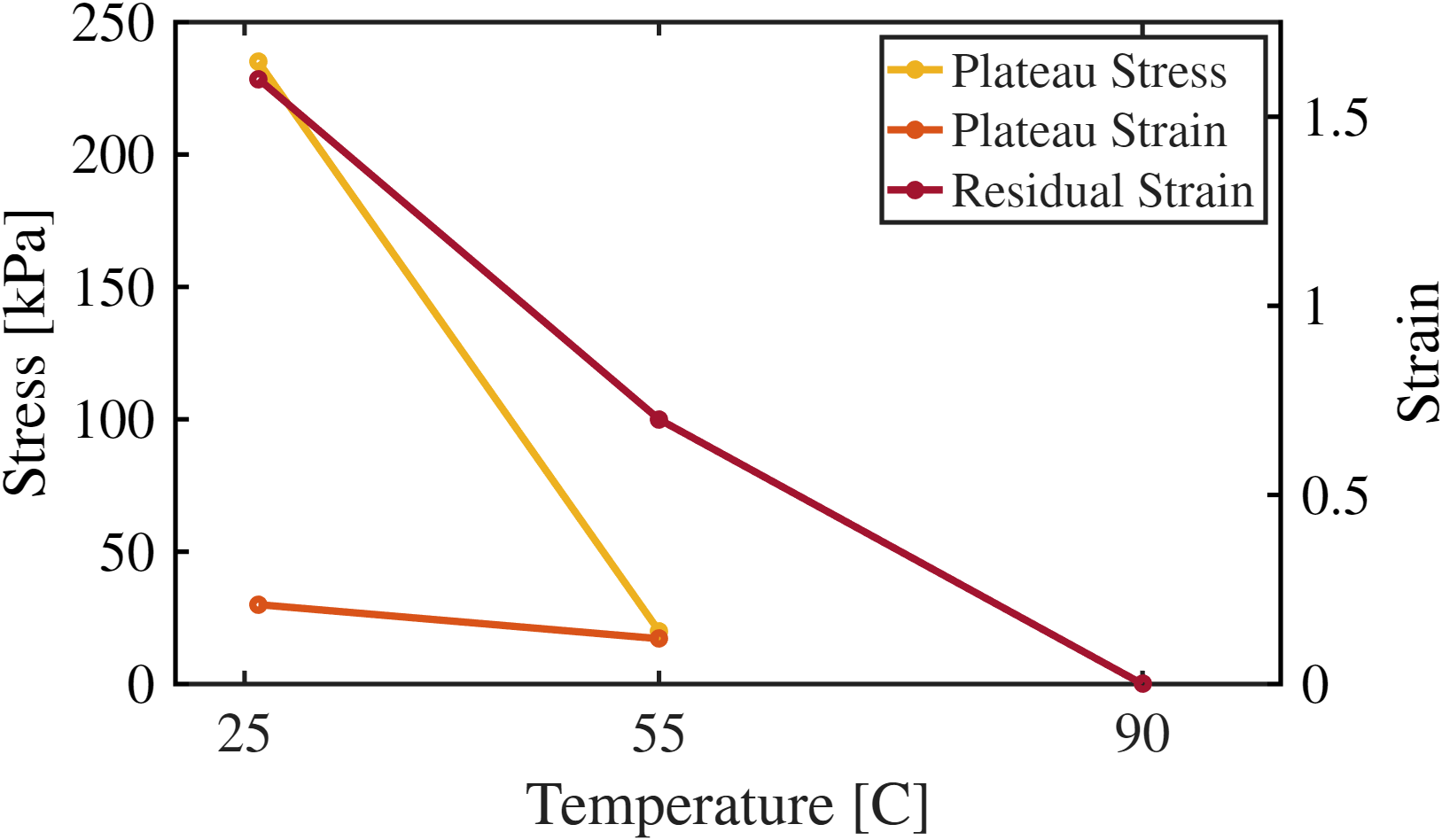}
     \caption{} \label{fig:plat_summ}
  \end{subfigure}
  \caption{Summary of temperature dependence at 5x10$^{-2}$ s$^{-1}$ nominal strain rate.  (a) Failure and (b) Soft behavior. Note that data points above the phase transition temperature are not included, since PMT regime does not exist.} 
  \label{fig:summ}
\end{figure}

The key aspects of the temperature dependence at a strain rate of 5$\times 10^{-2}$  s$^{-1}$ are summarized in Figure \ref{fig:summ}.  We observe that the stress and strain at failure decrease with increasing temperature.  This suggests that mesogen reorientation provides some toughening as suggested by \cite{kupfer_finkelmann, clarke_terentjev, tokumoto_zhou}. We also observe that the plateau and residual strain decrease with increasing temperature, consistent with the decrease of nematic order with increasing temperature. The plateau stress similarly decreases, suggesting that with increasing temperature, the network elasticity likely overshadowed the liquid crystal behavior, resulting in a softer response overall as temperature increases. This finding is consistent with \cite{azoug_2016, linares_2020, hotta_terentjev}, where it was also suggested that the network behavior is more dominant than the mesogen ordering at high temperature due to increased anisotropy, following the proposal of \cite{fridrikh_terentjev_1999}.

\subsection{Cross-linker density 25 mol\%} \label{sec:25mol}

\begin{figure}
   \centering
   \begin{subfigure}{0.48\textwidth}
     \includegraphics[width=\linewidth]{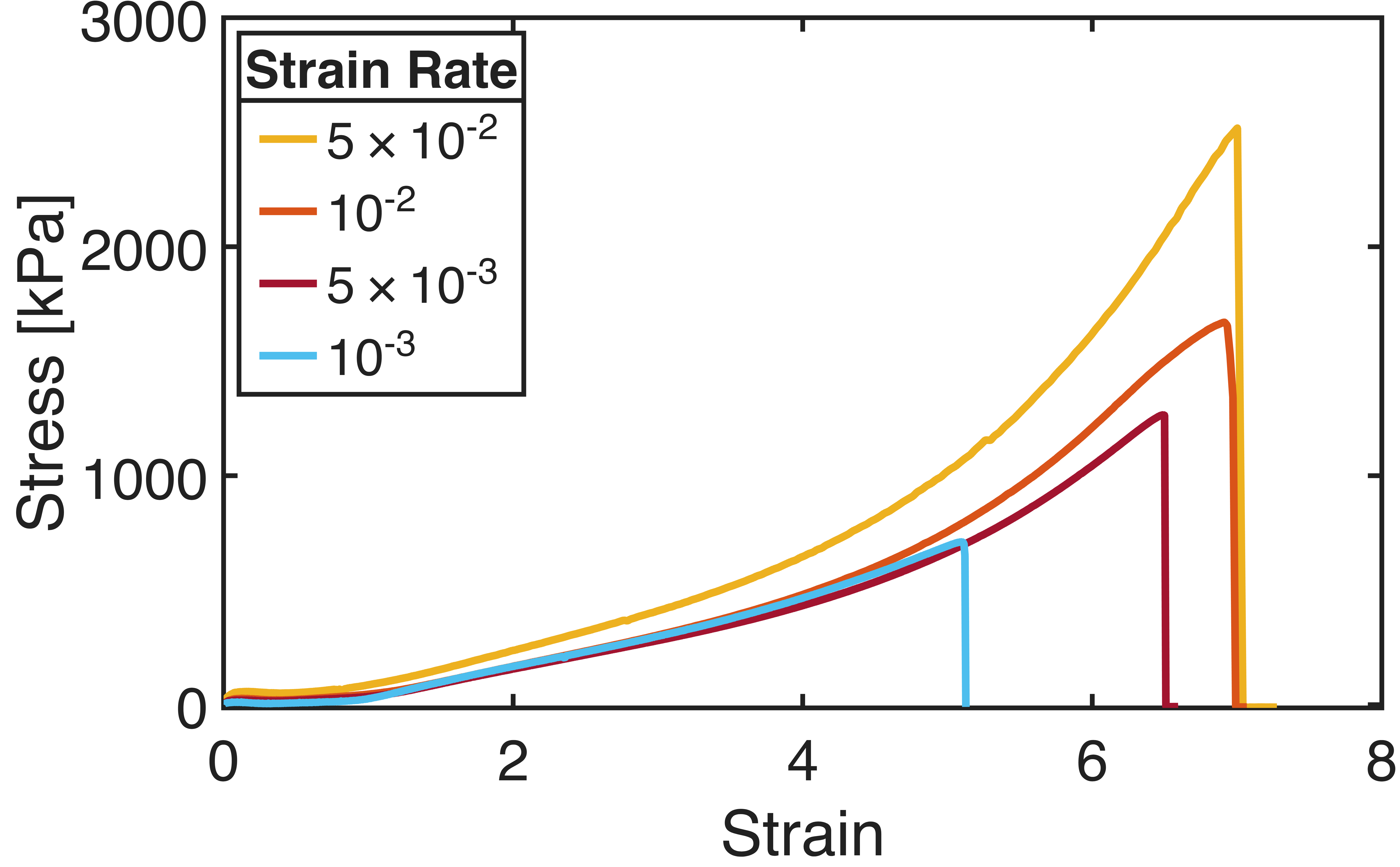}
     \caption{} \label{fig:25_a}
  \end{subfigure}%
  \hspace{0.01\textwidth}
  \begin{subfigure}{0.48\textwidth}
     \includegraphics[width=\linewidth]{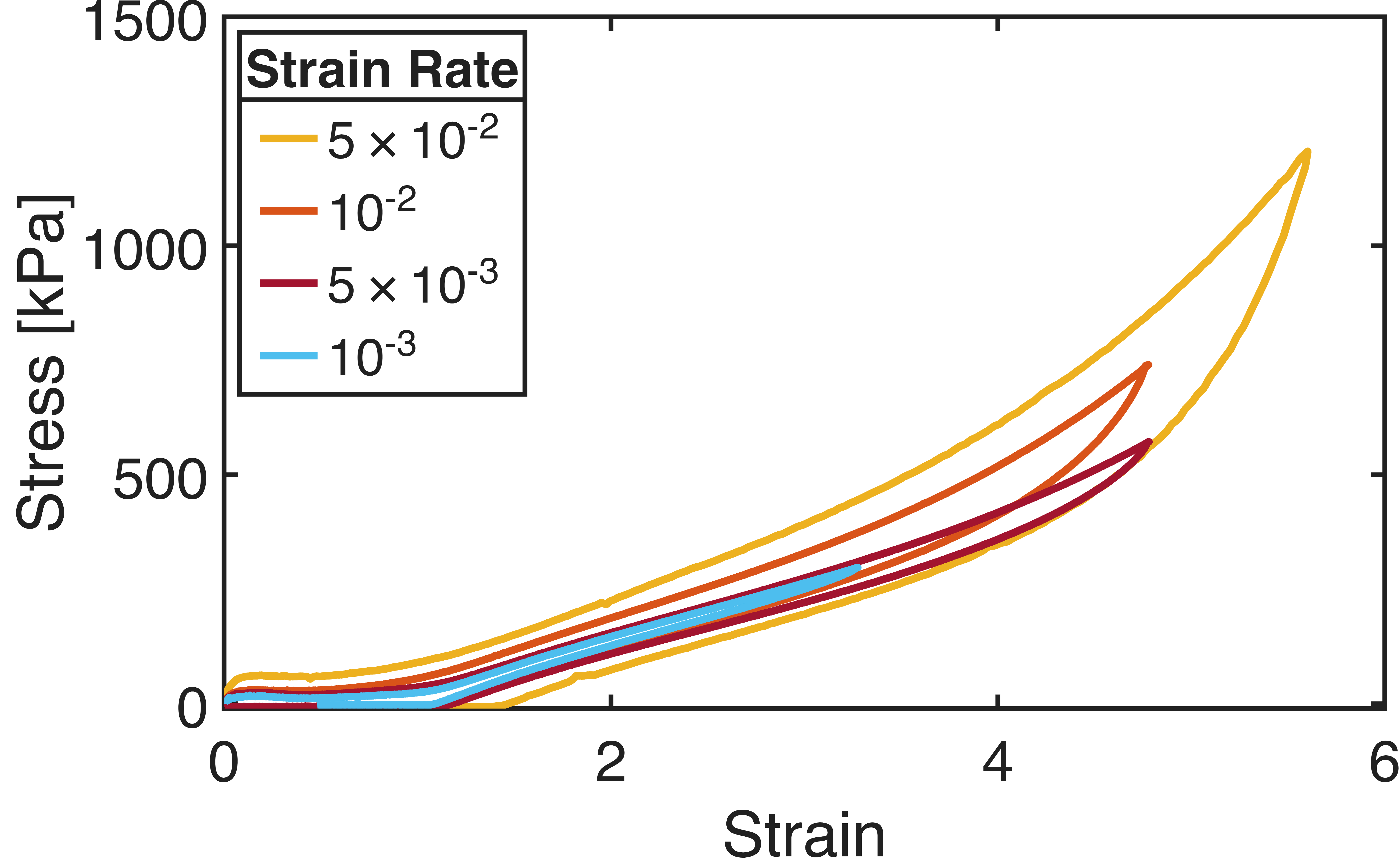}
     \caption{} \label{fig:25_b}
  \end{subfigure}%
  \hspace{0.01\textwidth}
  \begin{subfigure}{0.48\textwidth}
     \includegraphics[width=\linewidth]{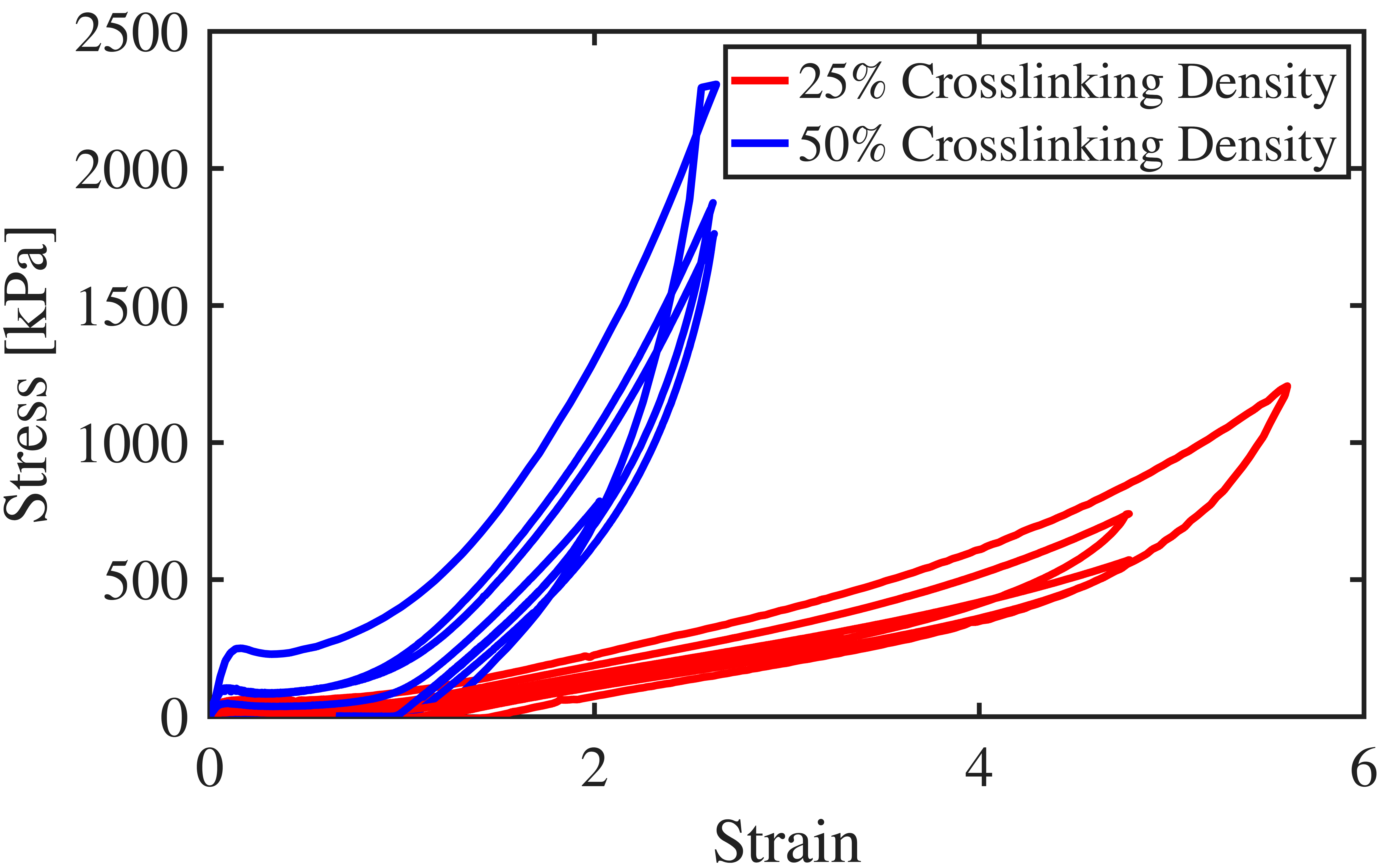}
     \caption{} \label{fig:25_c}
  \end{subfigure}%
\caption{Room temperature tensile for 25 mol\% cross-linker density at various strain rates and comparison. (a) Test to failure, (b) load-unload tests, and (c) comparison between the 25 mol\% and 50 mol\% cross-linker densities showing all four loading rates}. \label{fig:25}
\end{figure}

We perform similar tests at room temperature on I-PLCEs with a lower cross-linker density of 25 mol\%. These samples show significantly softer behavior, displayed in Figure \ref{fig:25}. DIC was not performed due to the high strains that result in large distortions of the speckles, and the data shown are based on the data output by the linear stage. The results are qualitatively similar to those described earlier for the 50 mol\% specimens, though these specimens are much softer and elongate to much greater strains before failure (for example, 7 vs 3.5 strain at a strain rate of 5x10$^{-2}$  s$^{-1}$). Figure \ref{fig:25_c} compares the loading curves at 25 mol \% and 50 mol \% across all strain rates, showing that the difference in crosslinking density significantly impacts plateau stress and strain relative to loading rate. Additionally, the end of the soft behavior or plateau regime is less obvious at lower cross-linker density, as the curve gradually increases rather than having a defined transition region. These results are consistent with previous studies, in which I-PLCEs with various cross-link densities (15 mol\%, 25 mol\%, 45 mol\%) were studied, and similar trends were observed \cite{Traugutt}. Since the PMT regime is present in both cross-link densities, we conclude that the trends across strain rates and temperatures we studied at higher cross-link densities should remain consistent at low cross-link densities.

\section{Fit to a macroscale constitutive model} \label{sec:model}
We now use the results to validate a model proposed in Lee {\it et al.} \cite{lee_2023}. Briefly, the model incorporates the evolution of microstructure (mesogens) through the internal variable framework of continuum mechanics. These internal variables encode the statistics of the mesogens at the microscopic scale and they capture the microstructure dynamics of the mesogens and domains from a coarse-grained perspective at the macroscopic scale. The model has successfully captured the behavior of I-PLCEs under multi-axial loading conditions observed experimentally in biaxial tests \cite{tokumoto_zhou}, Hertzian contact \cite{farre2022dynamic,ohzono2019enhanced}, and adhesion \cite{maghsoodi_2025} in the quasi-static setting. In the dynamic setting, it has been extended to incorporate network viscosity and has successfully captured the behavior of I-PLCEs under impact loading \cite{wihardja2025high}. 

To capture microstructure dynamics in the macroscopic scale, the model introduces two internal variables $(\Lambda,\Delta)$ that describe the domain patterns:  $\Lambda$ serves as a measure of the degree of alignment of all mesogens along the principal stretch direction in any representative volume element (RVE) while $\Delta$ serves as a measure of the degree of planarity of all mesogens in the RVE.  The model assumes isotropy and incompressibility, and thus the deformation is characterized by the principal stretch $\lambda$ and the product of the two largest principal stretches $\delta$.  The stored energy density is postulated to be 
\begin{align}
    W = \frac{\mu}{2} \left(\frac{\lambda^2}{\Lambda^2} + \frac{\delta^2\Lambda^2}{\lambda^2\Delta^2} + \frac{\Delta^2}{\delta^2} -3 \right) + C\frac{\Delta-1}{(r^{1/6}-\Delta)^k}
\end{align}
where $r$ is the so-called anisotropy parameter that describes the degree of nematic order, $\mu$ is the shear modulus, and $C$ is a hardening coefficient, and $k$ is a hardening exponent.  The first term in the energy is the elastic energy of the polymer network accounting for the spontaneous deformation of the domains, while the second term describes the memory of the domains and their energetic cost to align, as well as the saturation of the internal variables at perfect alignment state. In particular, this contribution is small near the isotropic case but increases as domain patterns evolves to be fully ordered in the plane, due to incompatibility from neighboring domains. Given the energy form above, the Piola-Kirchhoff stress is
\begin{align}
S = - p I + \frac{\partial W}{\partial F}
\end{align}
where $p$ is the constitutively indeterminate hydrostatic pressure.  The evolution laws of the internal variables are taken to be linearly dependent on their thermodynamic driving forces, hence
\begin{equation} \label{eq:evol}
\dot \Lambda = - \alpha_\Lambda \frac{\partial W}{\partial \Lambda}, \quad \dot \Delta = - \alpha_\Delta \frac{\partial W}{\partial \Delta}.
\end{equation}
Further details are found in \cite{lee_2023}.

We note that the model focusses on domain reorientation, and therefore ignores network viscosity.  This is consistent with our experiments with high crosslink density.  Still, the model is rate-dependent through these evolution equations \ref{eq:evol}. We refer the reader to \cite{wihardja2025high} for the extension of this model to include network viscosity and high strain rates.  We also note that the model is isothermal, though the parameters depend on temperature.

\begin{figure}
   \centering
   \begin{subfigure}{0.31\textwidth}
     \includegraphics[width=\linewidth]{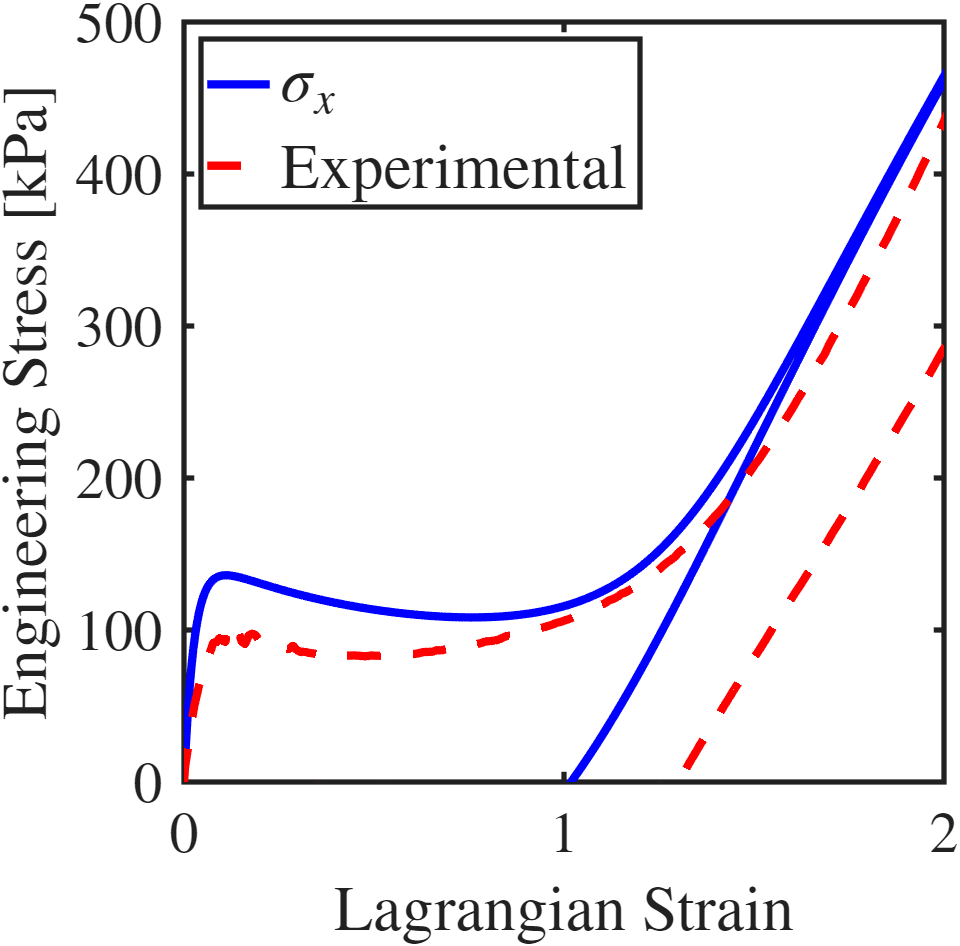}
     \caption{} \label{fig:Versus_Model_a}
  \end{subfigure}%
  \hspace{0.02\textwidth}
  \begin{subfigure}{0.31\textwidth}
     \includegraphics[width=\linewidth]{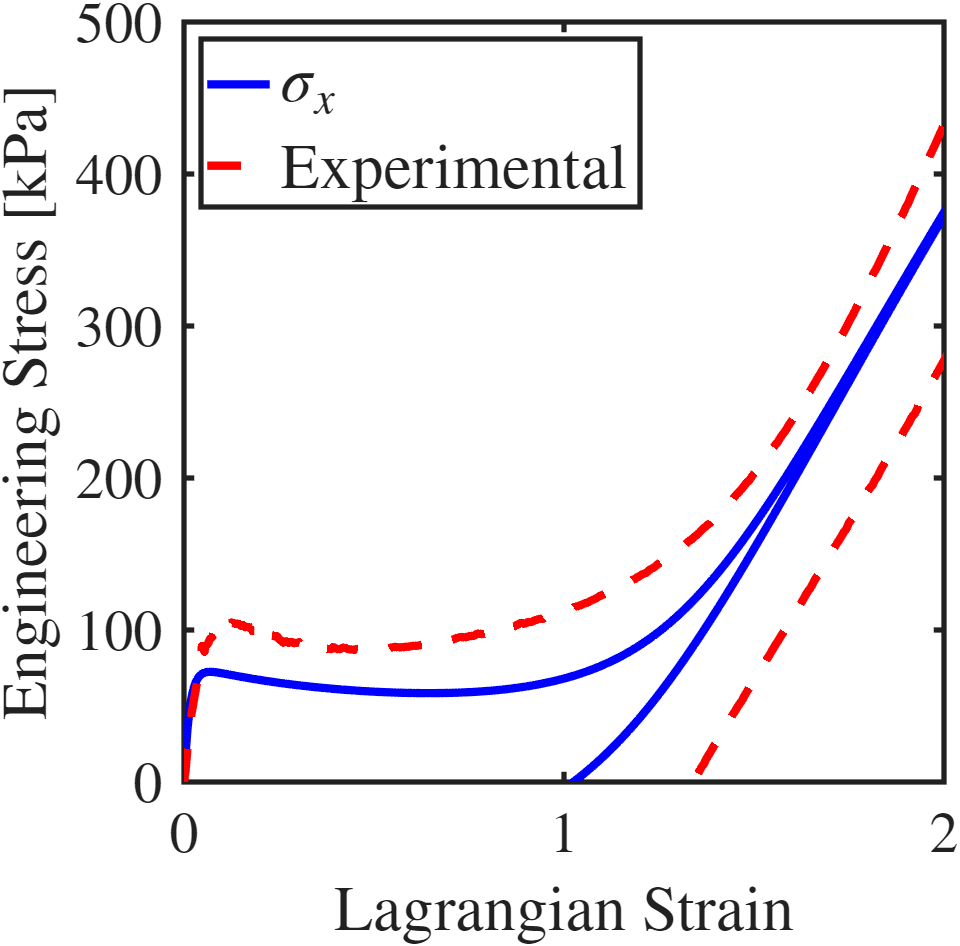}
     \caption{} \label{fig:Versus_Model_b}
  \end{subfigure}%
  \hspace{0.02\textwidth}
  \begin{subfigure}{0.31\textwidth}
     \includegraphics[width=\linewidth]{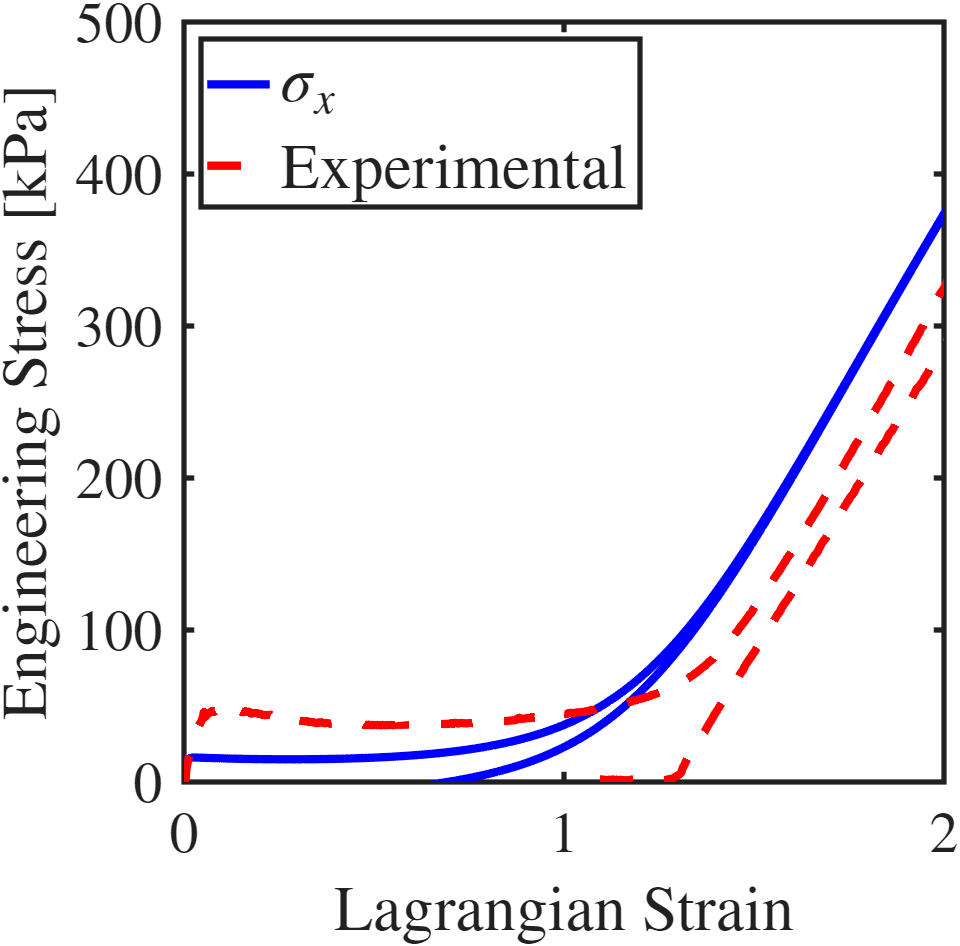}
     \caption{} \label{fig:Versus_Model_c}
  \end{subfigure}%
  \hspace{0.1\textwidth}
  \begin{subfigure}{0.31\textwidth}
     \includegraphics[width=\linewidth]{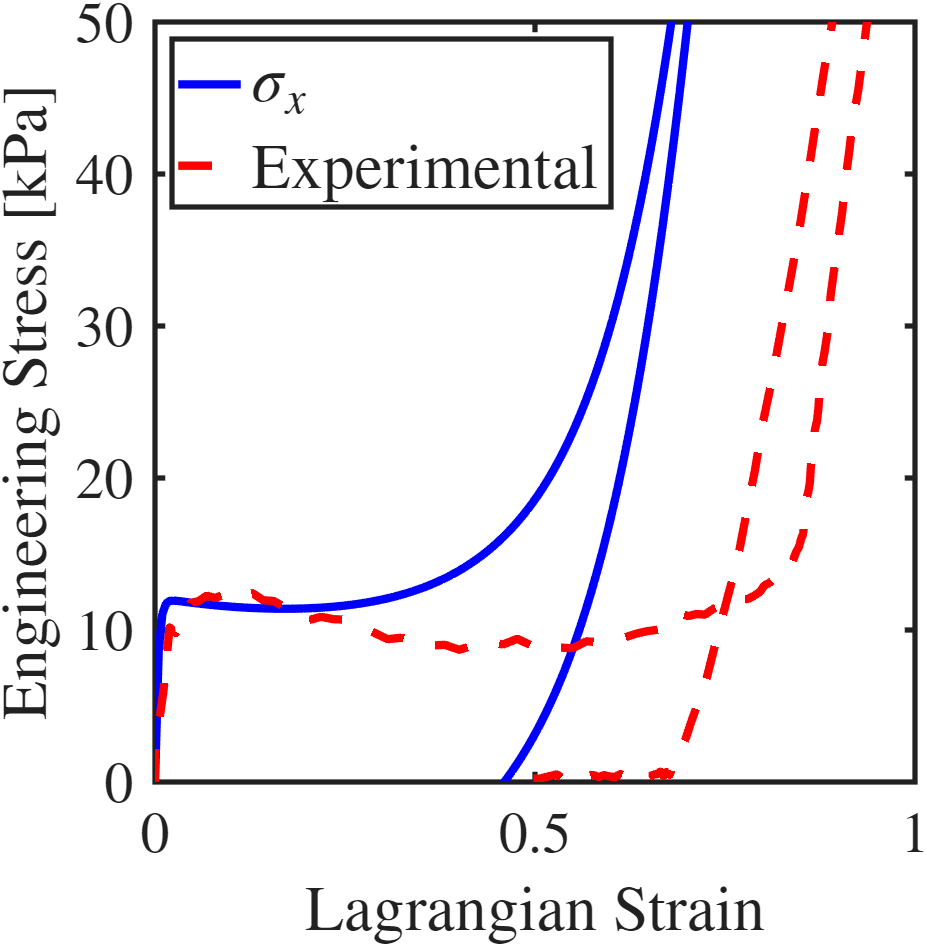}
     \caption{} \label{fig:Versus_Model_d}
  \end{subfigure}%
  \hspace{0.02\textwidth}
  \begin{subfigure}{0.31\textwidth}
     \includegraphics[width=\linewidth]{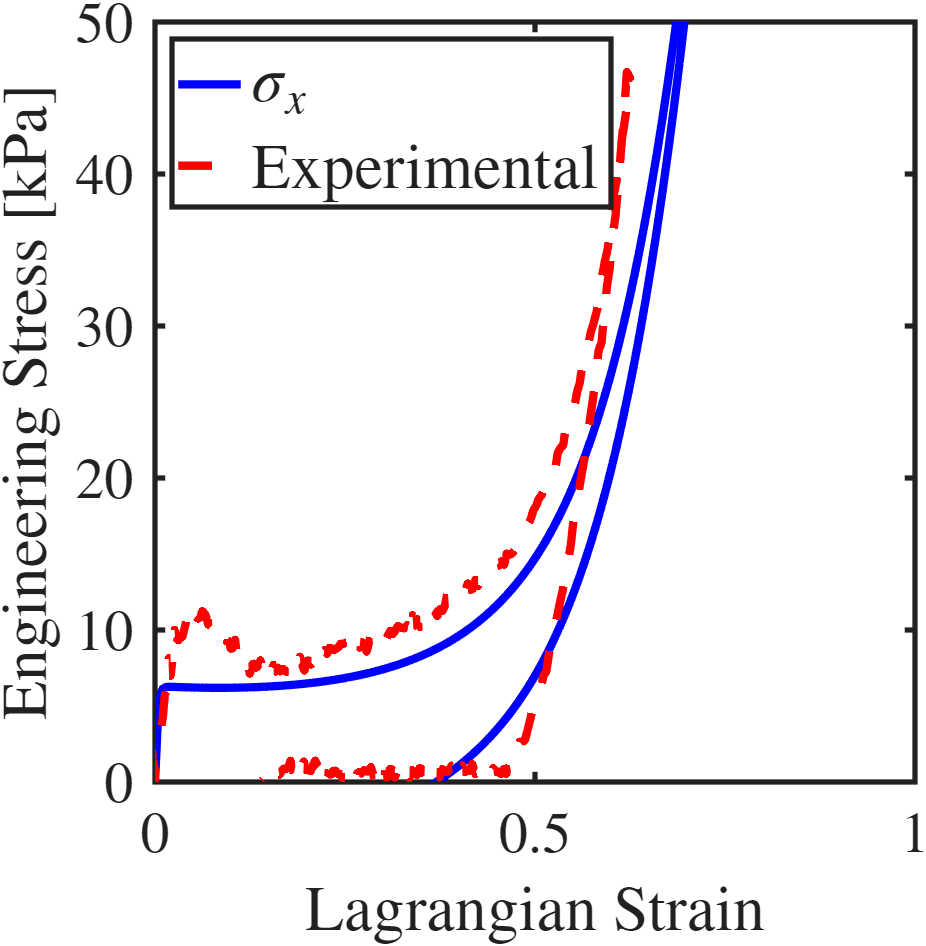}
     \caption{} \label{fig:Versus_Model_e}
  \end{subfigure}%
  \hspace{0.1\textwidth}
  \begin{subfigure}{0.31\textwidth}
     \includegraphics[width=\linewidth]{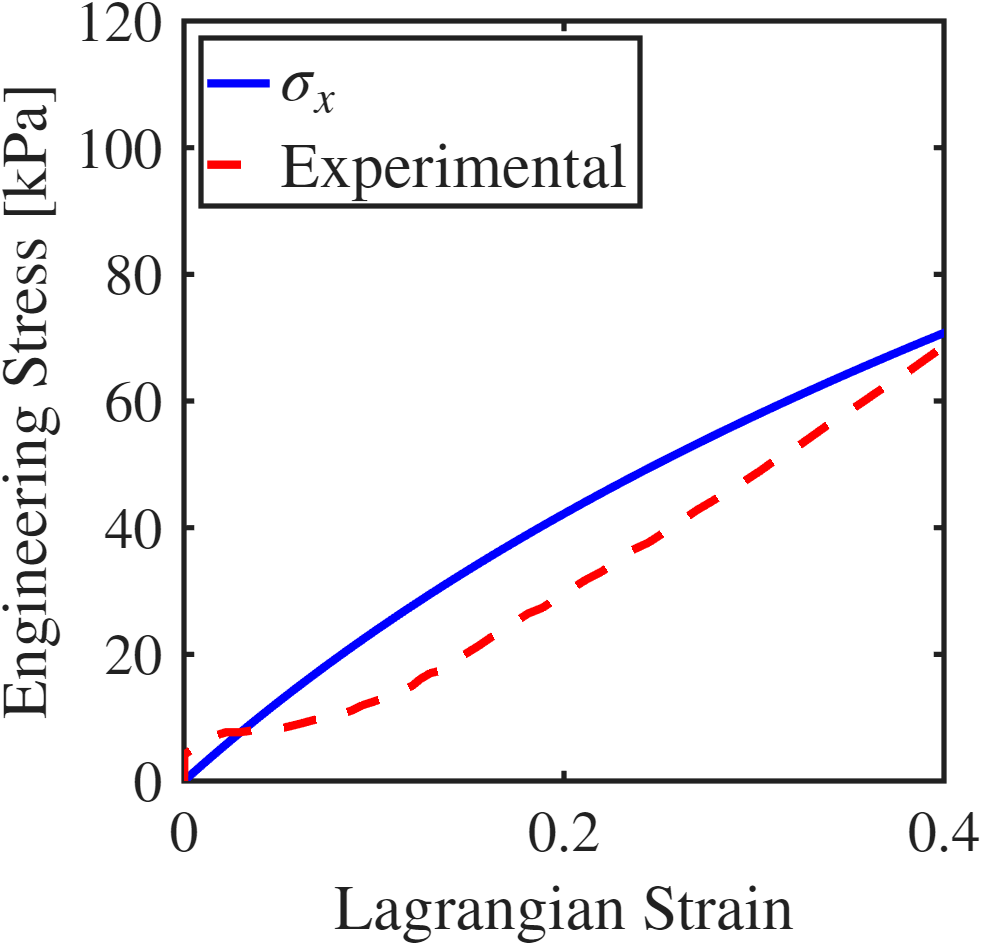}
     \caption{} \label{fig:Versus_Model_f}
  \end{subfigure}%
  \hspace{0.02\textwidth}
  \begin{subfigure}{0.31\textwidth}
     \includegraphics[width=\linewidth]{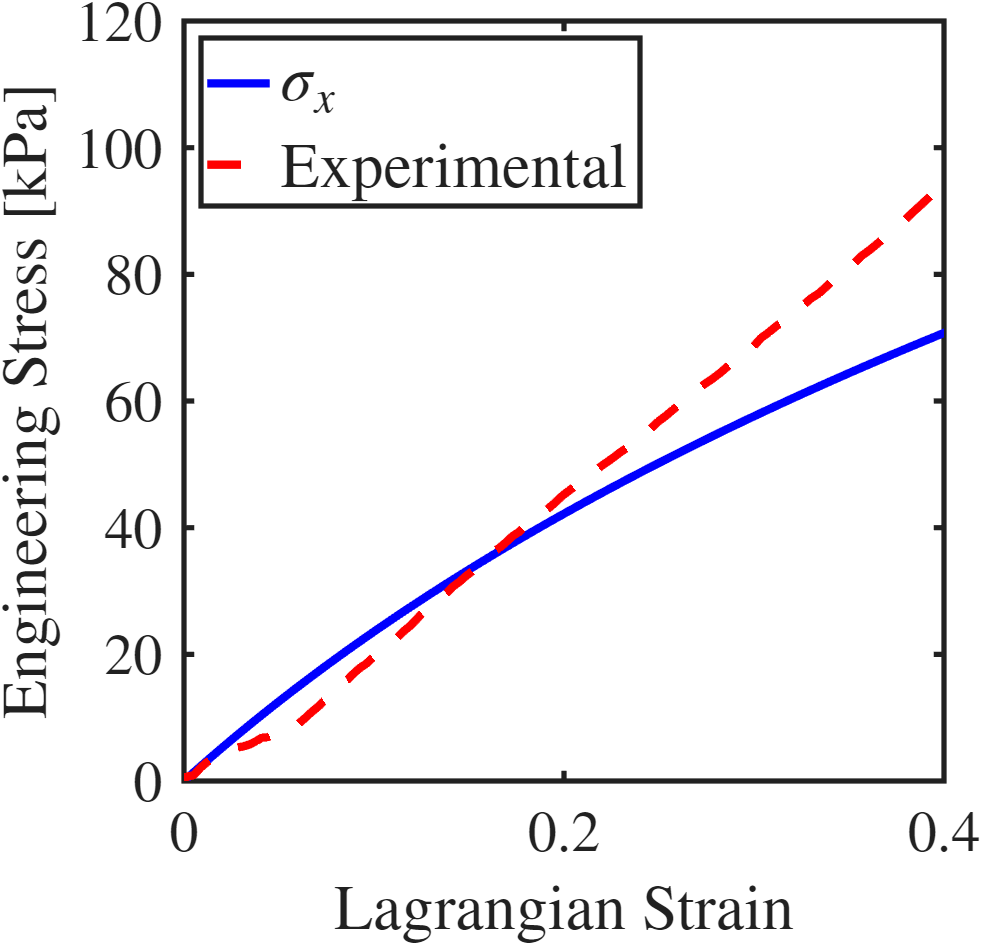}
     \caption{} \label{fig:Versus_Model_g}
  \end{subfigure}%
  \hspace{0.02\textwidth}
  \begin{subfigure}{0.31\textwidth}
     \includegraphics[width=\linewidth]{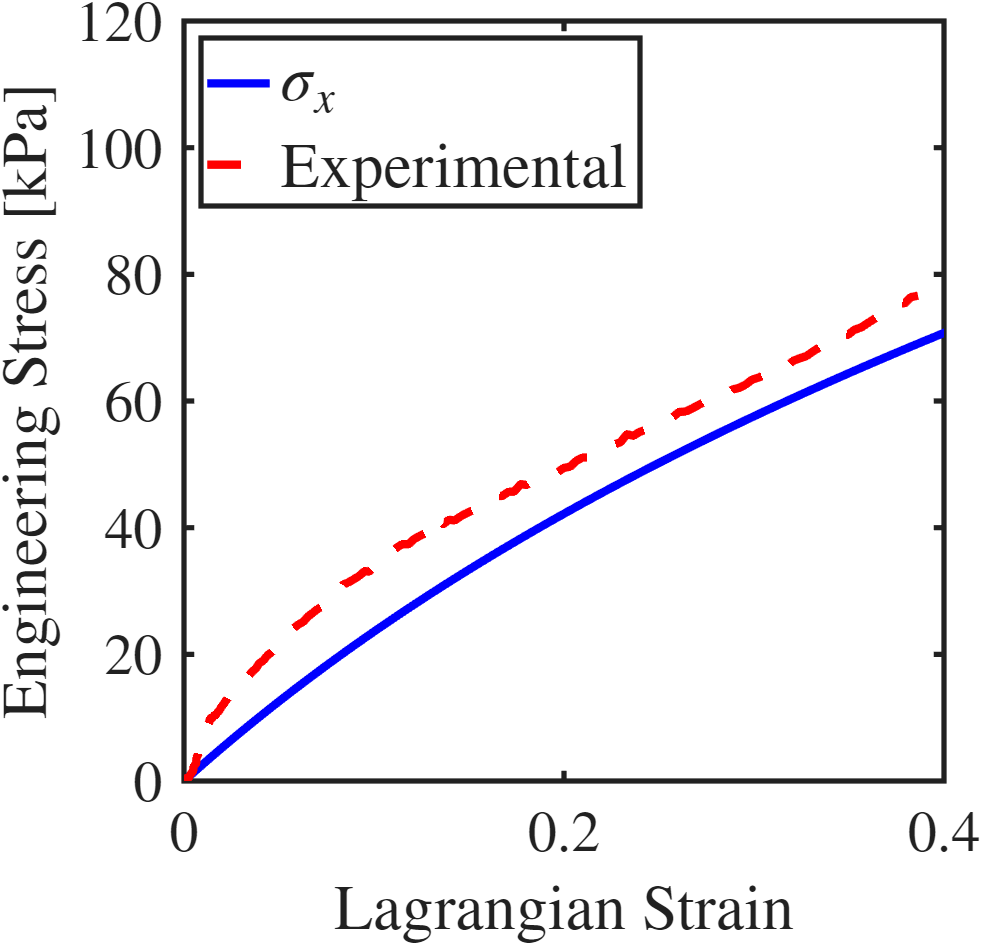}
     \caption{} \label{fig:Versus_Model_h}
  \end{subfigure}%
\caption{ Experimental data compared to the model, for each parameter set, listed as temperature, strain rate: (a) RT, $10^{-2}$ s$^{-1}$, (b) RT, 5$\times10^{-3}$ s$^{-1}$, (c) RT, $10^{-3}$ s$^{-1}$, (d) 55\si{\degreeCelsius}, $10^{-2}$ s$^{-1}$, (e) 55\si{\degreeCelsius}, 5$\times10^{-3}$ s$^{-1}$, (f) 90\si{\degreeCelsius}, $10^{-2}$ s$^{-1}$, (g) 90\si{\degreeCelsius}, 5$\times10^{-3}$ s$^{-1}$, and (h) 90\si{\degreeCelsius}, $10^{-3}$ s$^{-1}$.} \label{fig:Versus_Model}
\end{figure}

The loading portion of one stress-strain curve from each temperature and strain rate pair at 50 mol\% cross-linker density is used to fit the model.    The resulting parameters are shown in Table \ref{tab:parameters}, and the fits in  Figure \ref{fig:Versus_Model}.  Note that all parameters are held fixed across strain rates at each temperature.   Given that much of the elastic and nematic behavior is entropic in nature, it is natural to expect temperature dependence.  We observe in Figure \ref{fig:Versus_Model} that the model describes the behavior extremely well.

We emphasize that we only use the loading portion of the experiments to fit the model.  So the unloading behavior serves as an independent test, and shows good agreement.  We observe some offset in the unloading strains, and there are a few reasons.  Recall that some of tests had loops during the loading to unloading transition.  Further, we observed occasional cracks near the grips at large stretch and also some slippage at the grips.  Finally, our tests go to very large strains, well beyond the Gaussian chain assumption of the elasticity in our model.  We believe that these are the reasons for the off-set in the experimental observations.

%
%
%

\begin{table}
    \centering
    \setlength{\extrarowheight}{2pt}
    \caption{Temperature dependent parameters for each model curve shown in Figure \ref{fig:Versus_Model}.}
    \label{tab:parameters}
    \begin{tabular}{m{2cm}|c| c| c }
        \hline
        Parameter &Room Temperature & 55\si{\degreeCelsius} & 90\si{\degreeCelsius} \\ \hline
        r & 10 & & 1  \\
        $\mu$ (kPa) & 2000 & 2000 & 90 \\
        k & 2 & 2 & 2  \\
        $\alpha_{\Delta}$ (Pa) & 6$\times 10^{4}$  & 4.5$\times 10^{3}$ & 1 \\ 
        $\alpha_{\Lambda}$ (Pa) & $\alpha_{\Delta}$/100 & $\alpha_{\Delta}$/100 & $\alpha_{\Delta}$/100  \\
        C (Pa) & 0.4  & 0.09 & 0.01 \\ \hline 
    \end{tabular}
\end{table}
%

\section{Conclusion} \label{sec:conclusion}

In this study, we have used a new tensile-testing experimental platform to probe the soft-elasticity behavior in isotropic-genesis polydomain liquid crystal elastomers (I-PLCEs) over a wide range of temperatures (room temperature to beyond the phase-transition temperature) and across various strain rates ($10^{-3}$ to $5 \times 10^{-2}$ s$^{-1}$) and cross-linker densities (25\% and 50\%). Below the nematic-isotropic phase transition, all specimens exhibited the characteristic stress plateau of I-PLCEs, with both plateau stress and strain and residual strain decreasing as temperature increased and strain rate decreased. Above this transition temperature, the plateau regime vanished, and there was no noticeable strain-rate–dependent response of the polymer network. Moreover, reducing the cross-linker density resulted in softer behavior and higher failure strains. We further fit a continuum model that incorporates the microstructural kinetics of the domain patterns to our experiments, showing excellent agreement across all studied temperatures and strain-rate regimes using a single temperature-dependent parameter set. Our experimental results elucidate the coupled effects of temperature, strain rate, and network density on the soft-elastic behavior of I-PLCEs. Furthermore, our model is shown to capture these coupled effects and can serve as a predictive framework for designing I-PLCEs for various applications. 

We conclude by noting that while we have probed the material at various temperatures, this work is limited to isothermal applications.  It would be of interest to develop a complete thermo-mechanical model that combines systematic calorimetry with mechanical tests at more temperatures.  This remains a task for the future.

\paragraph{Conflict of interests.} The authors declare that they have no conflict of interests.

\paragraph{Acknowledgements.} The authors gratefully acknowledge the financial support of the US National Science Foundation (DMS-2009289) and the US Army Research Office (W911NF-22-1-0269).

\newpage
\bibliographystyle{abbrv} %
\bibliography{refs}

\section*{Appendix}
\appendix
\section{Experimental Setup Details} \label{sec:experimental details}
Various parts used in the experimental setup, organized by subsystem, are:
\begin{itemize}
   \item Custom-made chamber assembly
   \begin{itemize}
     \item Chamber components (McMaster 1658A12, 8983K128, 8983K118, 2313N23)
     \item Various hardware (nuts, bolts, washers, etc.)
     \item Various adapters (McMaster 8489K44, 4416T47)
     \item Windows and sealing (McMaster 8476K999, 92320A662)
     \item Stationary bottom clamp and moving clamp on pullrod (McMaster 12755T82)
     \item Moving pullrod (McMaster 8489K46)
   \end{itemize}
   \item Heating
   \begin{itemize}
     \item Two heaters (Omega OTF-102/120V)
     \item Two RTD air temperature sensors (Omega RTD-805-B)
     \item Temperature controller (Omega CS8DPT)
     \item 3M High-Temperature Flue Tape
   \end{itemize}
   \item Extension
   \begin{itemize}
     \item Linear stage (Physik Instrumente M-531.EC)
     \item Linear stage controller (Physik Instrumente C-863.11)
     \item Linear stage z-axis mounting bracket (Physik Instrumente M-592.10)
     \item Suspension system (ThorLabs VB01B)
   \end{itemize}
   \item Load
   \begin{itemize}
     \item Load cell (Omega LC101-50)
     \item External 10V power supply
   \end{itemize}
   \item Optics
   \begin{itemize}
     \item Vibration-isolation table (ThorLabs T46J)
    \item Continuous light source
     \item Camera (NOVA-S12\-8M)
     \item Photron FASTCAM Viewer software
   \end{itemize}
   \item Data acquisition
   \begin{itemize}
     \item Lab computer (Lenovo ThinkStation P330)
     \item DAQ (Omega INET-600) 
     \item MATLAB software
     \item VIC-2D software
   \end{itemize}
 \end{itemize}

Since tests are performed at various temperatures up to 90\si{\degreeCelsius}, the pullrod and one of the adapters were made out of ceramic to prevent the load cell from overheating. Binder clips were used to clamp the samples because they are self-tightening, and the inside of the clips were sanded to increase friction between the sample and the clip.

\section{Batch and temperature variation} \label{app:var}
\renewcommand{\thetable}{A\arabic{table}}
\setcounter{table}{0}

The average stress and standard variation across various batches tested at room temperature at a loading rate of $10^{-2}$ s$^{-1}$ is given in Table \ref{tab:batch_std}.  The mean stress at various temperatures tested at a loading rate of $10^{-2}$ s$^{-1}$ is given in Table \ref{tab:temp_values}.

\begin{table}
    \centering
    \caption{Average stress and standard deviation at various strains across batches at room temperature at a loading rate of $ 10^{-2}$ s$^{-1}$.}
    \label{tab:batch_std}
    \renewcommand{\arraystretch}{1.25}
    \begin{tabular}[t]{ C{3em} C{8em} C{8em} } 
        \hline
         Strain & $\sigma_{avg}$ (kPa) & $\sigma_{std}$ (kPa) \\ 
        \hline
        0.25 & 104.02 & 24.66 \\ 
        \hline
        0.5 & 137.67 & 53.64 \\ 
        \hline
        1 & 228.26 & 17.97 \\ 
        \hline
        1.5 & 540.90 & 65.05 \\ 
        \hline
        2 & 1021.74 & 134.44 \\ 
        \hline
        2.5 & 1689.20 & 236.41 \\ 
        \hline
    \end{tabular}
\end{table}

\begin{table}
    \centering
    \caption{Direct comparison of stress values at specific strains across temperatures. All data is taken from trials with a strain rate of $10^{-2}$ s$^{-1}$.}
    \label{tab:temp_values}
    \renewcommand{\arraystretch}{1.25}
    \begin{tabular}[t]{ C{3em} C{8em} C{9em} C{9em} } 
        \hline
         Strain & $\sigma_{avg}$ at RT (kPa) & $\sigma_{avg}$ at 55\si{\degreeCelsius} (kPa) & $\sigma_{avg}$ at 90\si{\degreeCelsius} (kPa) \\ 
        \hline
        0.25 & 104.02 & 10.67 & 86.40 \\ 
        \hline
        0.5 & 137.67 & 8.98 & 166.20 \\ 
        \hline
        1 & 228.26 & 74.81 &  \\ 
        \hline
    \end{tabular}
\end{table}

\end{document}